\documentclass[prd,reprint,superscriptaddress,nofootinbib,nobalancelastpage]{revtex4-1}
\usepackage{amsmath}
\usepackage{amssymb}
\usepackage{amsfonts}
\usepackage{graphicx}
\usepackage[caption=false]{subfig}
\usepackage{enumitem}
\usepackage[usenames,dvipsnames]{color}
\usepackage{verbatim}
\usepackage[percent]{overpic}
\usepackage{bm}
\usepackage{rotating}
\usepackage{hyperref}
\usepackage{pbox}
\usepackage{array}
\usepackage{physics}
\usepackage{mathtools}
\usepackage{leftidx}
\usepackage{lmodern}
\usepackage[normalem]{ulem}
\usepackage{braket}
\usepackage{tensor}
\usepackage{mathrsfs}
\usepackage{float}
\usepackage{tikz}
\usetikzlibrary{arrows.meta,decorations.pathmorphing,math}

\newcommand{\detg}{\sqrt{-g}}
\newcommand{\pd}{\partial}
\newcommand{\R}{\mathbb R}
\newcommand{\C}{\mathbb C}
\newcommand{\Z}{\mathbb Z}
\newcommand{\rr}[1]{\left(#1\right)}
\newcommand{\com}[1]{\left[#1\right]}

\newcommand{\sx}{\mathsf{x}}

\definecolor{plum}{rgb}{0.36078, 0.20784, 0.4}
\definecolor{chameleon}{rgb}{0.30588, 0.60392, 0.023529}
\definecolor{cornflower}{rgb}{0.12549, 0.29020, 0.52941}
\definecolor{scarlet}{rgb}{0.937, 0.161, 0.161}
\definecolor{brick}{rgb}{0.64314, 0, 0}
\definecolor{sunrise}{rgb}{0.80784, 0.36078, 0}

\usepackage{suffix}
\DeclarePairedDelimiterX\MeijerM[3]{\lparen}{\rparen}%
{\begin{smallmatrix}#1 \\ #2\end{smallmatrix}\delimsize\vert\,#3}

\newcommand\MeijerG[8][]{%
	G^{\,#2,#3}_{#4,#5}\MeijerM[#1]{#6}{#7}{#8}}

\WithSuffix\newcommand\MeijerG*[7]{%
	G^{\,#1,#2}_{#3,#4}\MeijerM*{#5}{#6}{#7}}

\allowdisplaybreaks

\usepackage[export]{adjustbox}

\begin{document}
	
	\title{Particle Detectors, Cavities, and the  Weak Equivalence Principle}

	\author{Erickson Tjoa}\email{e2tjoa@uwaterloo.ca}
	\affiliation{Department of Physics and Astronomy, University of Waterloo, Waterloo, Ontario, N2L 3G1, Canada}
	\affiliation{Institute for Quantum Computing, University of Waterloo, Waterloo, Ontario, N2L 3G1, Canada}

	\author{Robert B. Mann}\email{rbmann@uwaterloo.ca}
	\affiliation{Department of Physics and Astronomy, University of Waterloo, Waterloo, Ontario, N2L 3G1, Canada}
	\affiliation{Institute for Quantum Computing, University of Waterloo, Waterloo, Ontario, N2L 3G1, Canada}

	\author{Eduardo Mart\'{i}n-Mart\'{i}nez}	\email{emartinmartinez@uwaterloo.ca}
	\affiliation{Department of Applied Mathematics, University of Waterloo, Waterloo, Ontario, N2L 3G1, Canada}
	\affiliation{Institute for Quantum Computing, University of Waterloo, Waterloo, Ontario, N2L 3G1, Canada}
	\affiliation{Perimeter Institute for Theoretical Physics, 31 Caroline St N, Waterloo, Ontario, N2L 2Y5, Canada}

    \begin{abstract}
    
We analyze a quantum version of the  weak equivalence principle, in which we compare the response of a static particle detector crossed by an accelerated cavity with the response of an accelerated detector crossing a static cavity in (1+1)-dimensional  flat spacetime. We show, for both massive and massless scalar fields, that the non-locality of the field is enough for the detector to distinguish the two scenarios.  We find this result holds  for  vacuum and  excited field states of different kinds  and we clarify the role of field mass in this setup. 
\end{abstract}
    	
    \maketitle
	

    \section{Introduction}
    The weak equivalence principle (WEP) has been one of the central tenets of  gravitational physics. It has a variety of formulations, but it asserts that the local effects of motion in a curved spacetime  cannot be distinguished from those of an accelerated observer in flat spacetime.  The proviso of locality eliminates measurable tidal forces (that would originate, for example, from a radially  convergent gravitational field) acting upon finite sized physical bodies.  It implies that the trajectories of bodies with negligible gravitational binding energy are independent of  their composition and structure, and depend only their  initial positions and velocities. 

    With the development of the Unruh-DeWitt (UDW) model in quantum field theory, WEP can be analyzed in the presence of quantum fields in contrast to the original classical formulation. While not equivalent to a full quantum version of the WEP, this approach provides an operational means of understanding  some important aspects of the WEP in a quantum context.
    In particular, since UDW detectors  capture fundamental features of the light-matter interaction for atomic systems  \cite{Pozas2016}, one can operationally study the WEP by asking if a free-falling detector in a stationary cavity in uniform gravitational field has a different response from that of a stationary detector surrounded by a free-falling cavity. This problem has recently been revisited in the context of moving mirrors \cite{Fulling:2018lez}.  
       
    Renewed effort has been expended in recent years towards
    reanalyzing the role of atomic detector models coupled to a real scalar field with regards to the connection to gravitational phenomena.  It has been  argued that non-inertiality can be distinguished locally by exploiting nonlocal correlations of the field \cite{Dragan:2011zz,Ahmadzadegan:2013iua,Ahmadzadegan:2014pva}, effectively providing an accelerometer. An analysis of the behaviour of a UDW detector in a static cavity indicated that QFT may provide a way of  distinguishing between flat-space acceleration and free-fall in the near-horizon regime \cite{Ahmadzadegan:2013iua}. More recently \cite{Scully:2017utk} atoms falling through a cavity near an event horizon, together with short-wavelength approximations, led to radiation that is Hawking-like as seen by observers at spatial infinity. Even more recently,  an analysis of a moving mirror in cavity  \cite{Fulling:2018lez} has been used to argue once and for all that a ``qualitative WEP"  should hold in a quantum-field theoretic setting, and emphasized the importance of the initial state of the field in determining radiation from a moving mirror. This investigation focused on mirrors lacking internal degrees of freedom, but nonetheless had the advantage of providing information about the stress-energy tensor in the cavity, which was apparently missed in the past.
    
    Here we will complement these recent studies by showing that atomic UDW detectors  also exhibit a qualitative WEP. In particular, we revisit the old problem of computing the response of a static detector surrounded by an accelerating cavity, and the response of an accelerating detector that is surrounded by a static cavity.   The key issue here is not the composition of the detector (the body), but rather of its quantum field (vacuum) environment.
We consider the response for various field states, including the (scalar) vacuum, excited Fock states, and also single-mode coherent field states. We find that the mass of the quantum scalar field does not enter into the response of the detector apart from providing a degradation in the transition amplitude and larger mode frequencies in the mode decomposition. This is a consequence of the fact that the conformal invariance of the massless Klein-Gordon equation is not a physical effect, and is to be distinguished from the \textit{conformal flatness} of the spacetime under consideration. For non-vacuum field states, we show how resonance can be used to amplify the transition probability  {via co-rotating terms} and demonstrate the irrelevance of the mass of the field in the physics  underlying WEP.
    
    Our paper is organized as follows.  In Section~\ref{sec:equivprinciple}, we revisit the formulation of WEP and clarify the contexts in which this work and others, in particular \cite{Fulling:2018lez,Scully:2017utk}, are performed. In Section~\ref{sec:setup} we provide the standard setup and generic expressions for  a UDW detector coupled to a Klein-Gordon field, without restriction to the vacuum state of the field. In Section~\ref{sec:detector} we consider an accelerating detector traversing the entire static cavity, noting the necessary changes if detector starts accelerating somewhere within the cavity. In Section~\ref{sec:cavity} we consider a static UDW detector that encounters an accelerating cavity, entering one end and leaving the other
    due to the motion of the cavity; we also note the necessary modifications if the trajectory of the detector is changed. In Section~\ref{sec:excited} we consider various non-vacuum field excitations and the role of resonance between atomic gap and excited cavity modes. In Section~\ref{sec:rate} we compute transition rate to better understand if the difference between two scenarios are not averaged out by the transition probability calculations.
    
    Throughout we adopt $c=\hbar=1$  so that the mass parameter $m$ has units of inverse length.
    
	
    \section{Weak equivalence principle revisited}
    \label{sec:equivprinciple}
	
\begin{figure}[htp]
        \centering
        \includegraphics{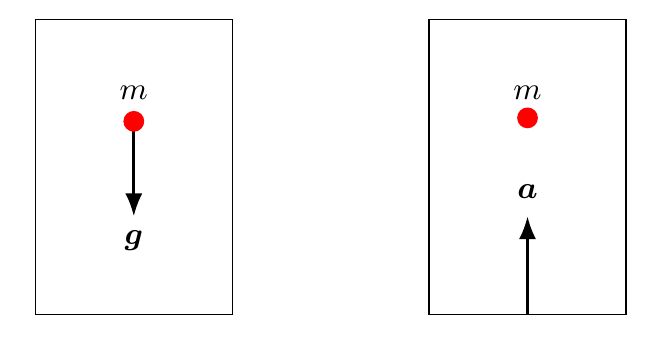}
	    \caption{Classical WEP setup: test mass $m$ in a closed cavity. WEP claims that these two are kinematically indistinguishable in absence of second-order effects such as tidal forces or non-uniform acceleration. The space inside the cavity is a true vacuum in the classical sense, apart from existence of gravitational field to mimic acceleration.}
	    \label{wep1}
    \end{figure}

	It is a remarkable property of gravity, in contrast to other non-gravitational forces, that every test particle equally and universally experiences the influence of gravitational fields.  This underlies the WEP, which states that  phenomenology of bodies   observed from frames in uniform gravitational fields is equivalent to that of frames that accelerate uniformly relative to inertial [free-falling] frames \cite{schutz2009first}. In other words, WEP states that a mass free-falling in a stationary cavity with uniform gravitational field $\bm g=-g\bm e_z$ is completely equivalent to a stationary mass with uniformly accelerating cavity\footnote{Alternatively, the normal force experienced by a test mass on the floor of a closed cavity cannot be attributed to cavity acceleration or uniform gravitational field without additional non-local information (e.g. by looking out of the cavity).} with $\bm a = g\bm e_z$, as shown in Figure~\ref{wep1}. This principle has been verified to great accuracy through various experiments and effectively sets inertial mass and gravitational mass to be equal. 
	
\begin{figure}[htp]
        \centering
        \includegraphics{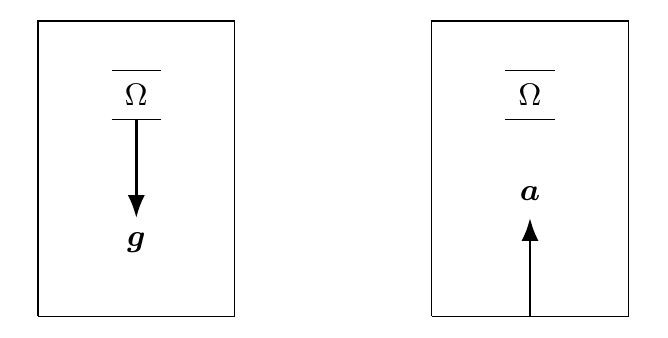}
	    \caption{Quantum WEP setup: a two-level atomic `detector' with gap $\Omega$ (with respect to its proper frame) in a cavity. The cavity contains a quantum field whose modes may lead to atomic excitations (even in the vacuum state), and hence detector responses. The atom serves as a detector in the sense that a particle is detected when the atom is excited and then emit radiation \cite{birrell1984quantum}.}
	    \label{wep2}
    \end{figure}

 The purely classical version of the WEP, while asserting that free-fall is independent of a body's composition,
    does not consider internal quantum degrees of freedom of a body (unlike a qubit).  For example, the body is considered to be uncharged and the space inside the cavity to be free of electromagnetic fields.	
	The quantum version of the WEP essentially requires us to consider an atomic detector coupled to some field prepared in some state, as shown in Figure~\ref{wep2}.   The state most closely resembling the classical environment of the WEP is the quantum vacuum.   Other field states  can of course be considered, but they will in general produce environments analogous to those of air or some other fluid that produces drag on the body.
        For example \cite{Fulling:2018lez}, having an electromagnetic field makes the argument less trivial: a classical electric charge in uniform accelerated motion radiates and it is nontrivial to ask whether a \textit{free-falling} charge radiates. The reason is because in general relativity, free-falling is an inertial motion (geodesic motion) and acceleration corresponds to non-geodesics in spacetime. According to the equivalence principle, however, we should be able to speak of uniform acceleration and constant gravitational field intercontrovertibly. Where is the problem?

  Here we revisit this problem using an Unruh-DeWitt (UDW) detector to replace the mirror in \cite{Fulling:2018lez}. Specifically, we consider two different ``Experiments":
    \begin{enumerate}[label=(\arabic*),leftmargin=*]
        \item \textbf{Stationary cavity, accelerating detector:} In this scenario, we let an atomic detector undergo uniform acceleration as it crosses the cavity containing the quantum field. This mimics the scenario of free-falling atom in gravitational field e.g. outside a black hole especially near the horizon, when the metric is approximately Rindler-like \cite{Scully:2017utk}.
        \item \textbf{Stationary detector, accelerating cavity:} In this scenario, a rigid cavity is accelerating such that a stationary detector traverses across the cavity. This mimics a free-falling cavity in gravitational field under appropriate quasilocal approximations.
    \end{enumerate}

    
 If WEP holds, then we should expect that Experiment 1 and Experiment 2 should be qualitatively symmetric in some regimes  for the scalar field vacuum. More generally., we might expect that for other field states that the motion
 of the body is independent of the field mass.  
  In order to make useful and valid comparisons in the context of WEP, we generally need to ensure two additional `requirements' on the setup in question. 
    
    First of all, we will need to be able to set up a kind of cavity undergoing constant acceleration  across its full spatial extent. This is forbidden in special relativity without abandoning the rigidity condition \cite{Herglotz1910,Noether1910,Epp:2008kk} (static boundary condition in Rindler coordinates). 
    Therefore, a rigid accelerating cavity suitable for WEP is necessarily   in a \textit{quasilocal} regime 
    in the sense of $aL\ll 1$ where $L$ is the proper length of the cavity at rest as measured in the lab frame. Outside this regime, we see that the accelerating cavity will have detectable non-uniform proper accelerations across the cavity and hence in the comoving frame of the cavity, the stationary detector (with respect to lab frame) does \textit{not} undergo uniform acceleration since the worldline of the detector crosses hypersurfaces of constant but different accelerations. Therefore, Experiment 1 and Experiment 2 are only equivalent in quasilocal approximations. 
    
    Secondly, we will need to show that the distinction between detector responses in Experiments 1 and 2 should be qualitatively independent of the mass of the quantum field and the initial state of the field within the quasilocal regime. In other words, \textit{quantitative} differences between Experiments 1 and 2 would then be attributed to nonlocal correlations: the atom is sensitive to the inequivalent setups in the two experiments and also the fact that  moving-boundary/stationary-atom is not the same as a moving-atom/stationary-boundary from a physical point of view. 
    
    Furthermore, the role of the field mass should only serve to degrade non-local correlations of the field and hence diminish transition amplitudes, all else being equal. This requirement, however, is in slight tension with previous results \cite{Dragan:2011zz} claiming (in the non-relativistic regime) that the field mass term can enhance the transition probability of a detector, making it a better accelerometer in the case of highly excited field states. This would mean that the mass of a scalar field leads to additional physical effects \textit{beyond} suppressing correlations. For the WEP  in particular,  one could imagine increasing the mass more and more to detect increasingly small local accelerations.  We will recover consistency with WEP by showing that this  discrepancy is in part due to mixing conformal flatness with conformal invariance of the Klein-Gordon equation.  We also note that the idea that massive excitations should be `harder' to detect than the massless ones, all things being equal, is not new --- it has been investigated e.g. in \cite{Kialka:2018a}. A more complete discussion of these issues is given in Appendix \ref{appendix:conformal} and \ref{appendix:disparity}.
    
    In light of these two requirements, in the next few sections we will consider the setup and demonstrate that the qualitative WEP is indeed observed. In particular, we recover the expectation that massless field should be able to detect relative acceleration (non-inertiality) as well as the massive field, if not better, in quasilocal regime.  The idea that detection of massive excitations should be `harder' than the corresponding massless ones is not new (see, for example, \cite{Kialka:2018a}). This entails the clarification that conformal invariance in the massless case has nothing to do with the physics of uniform acceleration and hence WEP; it is a computational convenience that one can invoke (cf. Appendix~\ref{appendix:conformal}), to be distinguished from the fact that all two-dimensional spacetimes are conformally flat. We will strengthen this claim by considering an arbitrary Fock state and a single-mode coherent state, and check that no  essential differences arise even in the transition rate (which is a differential version of the detector response).

    \section{Setup}\label{sec:setup}

    Our starting point is the Klein-Gordon equation for a real scalar field: the covariant formulation of Klein-Gordon equation which governs the dynamics of a real scalar field reads
    \begin{align}
    \label{kleingordon}
         \frac{1}{\detg}\pd_\mu{\left(g^{\mu\nu}\sqrt{-g}\pd_\nu\phi\right)} + m^2\phi = 0\,.
    \end{align}
    For global Minkowski spacetime, the solutions are given by plane waves. Recall that all $(1+1)$ dimensional spacetimes are conformally equivalent to Minkowski spacetime: by this we mean that \textit{there exists} a coordinate system in which the metric is conformally flat, i.e. with metric that takes the form
    \begin{equation}
        g_{\mu\nu}(\sx)=\Omega^2(\sx)\eta_{\mu\nu}\,.
    \end{equation}
    This conformal flatness can be exploited in the case of $m=0$ 
    to map the solutions of the Klein-Gordon equation to the plane-wave solutions in Minkowski spacetime because the massless Klein-Gordon equation is conformally invariant  in $(1+1)$ dimensions.
    This allows us to obtain an \textit{exact closed form} for the spectrum of the field modes\footnote{An important point here is that conformal invariance is convenient but not necessary. We show this in Appendix~\ref{appendix:conformal}.}. For $m\neq 0$, the conformal invariance of the wave equation is lost and hence conformal flatness provides no particular advantage. Therefore, even for a uniformly accelerating frame the field modes can be written in closed form;  however neither the normalization nor the spectrum can. 
    
    
    To probe the field, we consider a pointlike Unruh-DeWitt (UDW) detector whose interacting Hamiltonian is given by
    \begin{equation}
        \begin{aligned}
        \hat H_{I}(\tau) &= \lambda
        \chi(\tau)\hat\mu(\tau)\hat\phi(\tau,x(\tau))\,,\\
        \hat\mu    &= e^{i\Omega\tau}\hat\sigma^++e^{-i\Omega \tau}\hat\sigma^-\,,
    \end{aligned}
    \end{equation}
    where $\tau$ is the proper time of the detector, $\lambda$ is the coupling strength of the detector and the field, $\hat\mu(\tau)$ is the monopole moment of the detector, $\hat\sigma^\pm$ are $\mathfrak{su}(2)$ ladder operators characterizing the two-level atomic detector, and $\chi(\tau)$ is the switching function of the detector. Here $\hat\sigma^+\ket g = \ket e$ and $\hat \sigma^-\ket e = \ket g$ where $\ket g,\ket e$ refer to the ground and excited states of the atom respectively, separated by energy gap $\Omega$. Note that the interacting Hamiltonian is given in the Dirac picture.
    
    We consider the initial state to be a separable state $\ket{g}\otimes\ket{\psi} = \ket{g,\psi}$ where $\ket\psi$ is some initial pure state of the field. If the field is in some $\ket{\text{out}}$ state after the interaction and the detector is in excited state $\ket e$, then the transition probability of the detector is given by Born's rule after tracing out the field state:
    \begin{align}
        P(\Omega) = \sum_\text{out}\abs{\braket{e,\text{out}|\hat U|g,\psi}}^2
    \end{align}
    where the time evolution operator in the Dirac picture is
    \begin{align}
        \hat U = \mathcal T\exp\rr{-\frac{i}{\hbar}\int_{-\infty}^{\infty}\dd\tau\hat H_I(\tau)}\,.
    \end{align}
    Employing the Dyson expansion
    \begin{equation}
        \begin{aligned}
        \hat U &= \hat \openone + \hat U^{(1)} + O(\lambda^2)\,,\\
        \hat U^{(1)} &= -\frac{i}{\hbar}\int_{-\infty}^{\infty}\dd\tau\hat H_I(\tau)
    \end{aligned}
    \end{equation}
    we obtain the leading order contribution to the transition probability
    \begin{equation}
        \begin{aligned}
            P(\Omega) &= \lambda^2 \int\dd\tau\int\dd\tau'\chi(\tau)\chi(\tau')\times \\
            &\hspace{0.5cm} e^{-i\Omega(\tau-\tau')}W(\tau,\tau') + O(\lambda^4)\,,\\
            W(\tau,\tau') &= \braket{\psi|\hat\phi[\sx(\tau)]\hat\phi[\sx(\tau')]|\psi}\,,
        \end{aligned}
    \end{equation}
 where $W(\tau,\tau')$ is the pullback of the Wightman function on the detector's trajectory $\sx(\tau)$.  The remaining task is to compute the Wightman function for different scenarios and choose an appropriate switching function of the detector.

\begin{figure}
      \centering
      \includegraphics{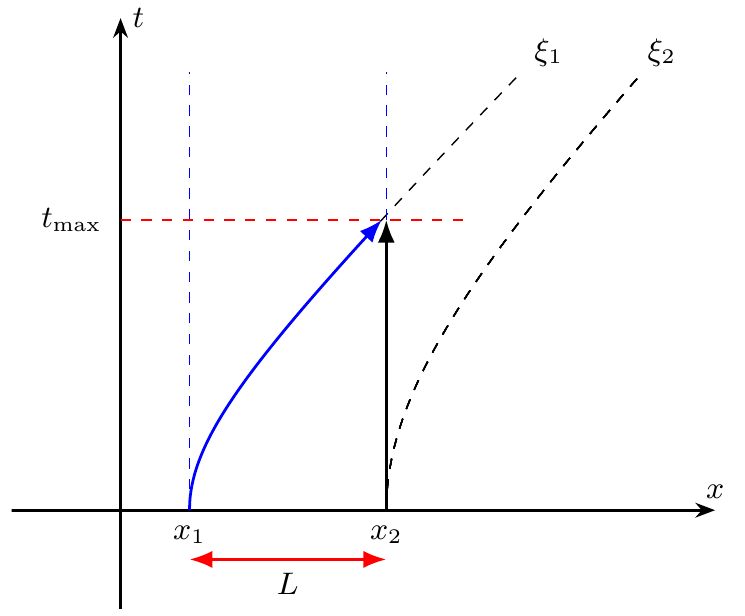}
      \caption{Spacetime diagram for the setup under consideration. The accelerating detector-static cavity scenario corresponds to the trajectories in blue, with accelerating detector denoted by the solid  curved arrow. The accelerating (rigid) cavity-static detector scenario corresponds to the trajectories in black, with the static detector denoted by the vertical solid vertical curve. In both scenarios the trajectories of the endpoints of the cavity are given by dashed lines.}
      \label{rindlerdiagram}
    \end{figure}

    We would like to study further the situation when one speaks of the weak equivalence principle in the presence of a quantum field subject to a  boundary condition 
    (a Dirichlet cavity) in (1+1) dimensions.
    We are interested in two types of scenarios (``Experiments") that can be summarized as follows:
    \begin{enumerate}[label=(\arabic*),leftmargin=*]
        \item A cavity is static relative to some laboratory frame $(t,x)$ and the detector is accelerating with constant proper acceleration. In the language of the equivalence principle, this should also describe a static cavity in a constant gravitational field (like on the surface of the Earth), with a free-falling detector.
        \item The detector is static relative to the lab frame and a \textit{rigid} cavity uniformly accelerates, mimicking a free-falling cavity in a uniform gravitational field.
    \end{enumerate}
    The spacetime diagram for these two setups are shown in Figure~\ref{rindlerdiagram}. Case (1) corresponds to the trajectories in blue, with accelerating detector denoted by solid curved arrow. Case (2) corresponds to trajectories in black, with static detector denoted vertical solid vertical curve. In both scenarios the cavity trajectories are in dashed lines.
    
    We will see that this result continues to hold even for excited field states, including   coherent states \cite{Simidzija:2017jpo}. Therefore, this effectively shows that massive scalar fields do not generally provide additional insight over massless scalar fields, apart from introducing an additional degradation factor  (for example  in the studies of entanglement generation or degradation \cite{Bruschi:2012}).

    \section{Accelerating detector}\label{sec:detector}
    Let us first consider the case of a static cavity relative to the lab frame $(t,x)$. In this case, the cavity is equivalent to a Dirichlet boundary condition $\phi(x_1)=\phi(x_2)=0$ where $x_j$ are the locations of the boundary and the length of the cavity as measured by the lab frame is $L=x_2-x_1$. The equation of motion for the quantized scalar field reduces to
    \begin{align}
        (\pd_\mu\pd^\mu+m^2)\hat\phi=0
    \end{align}
    and under a Dirichlet boundary condition the modes are standing waves:
    \begin{equation}
        \begin{aligned}
        \hat\phi(t,x) &= \sum_{n=1}^\infty u_n(x)\rr{e^{-i\omega_n t}\hat a_n+e^{i\omega_n t}\hat a_n^\dagger}\,,\\
        u_n (t,x) &= \frac{1}{\sqrt{L\omega_n}}\sin\frac{n\pi(x-x_1)}{L}\,,
    \end{aligned}
    \end{equation}
    where \mbox{$\omega_n^2 = \rr{\frac{n\pi}{L}}^2+m^2$}. The normalization of $u_n$ can be found using the Klein-Gordon inner product \cite{birrell1984quantum}
    \begin{equation}
    \label{KGinnerproduct}
        (\phi_1, \phi_2) = -i\int_\Sigma\dd\Sigma^\mu\rr{ \phi_1\pd_\mu\phi_2^* -(\pd_\mu\phi_1)\phi_2^*}\,,
    \end{equation}
    where $\Sigma$ is the spacelike hypersurface foliated by the global time function defining the timelike Killing vector of the spacetime.
    
    For the detector/cavity configuration (see figure \ref{rindlerdiagram}), if the detector is accelerating from left wall to the right wall of the cavity then $x_1=0$; if the detector starts from midpoint, then $x_1=-L/2$.  Starting at the midpoint (as  in \cite{Dragan:2011zz}) is useful if we wish to consider the $a=0$ limit, since the Dirichlet boundary renders the limit ill-defined for a detector starting from the left where the field vanishes  (i.e. the detector `merges' with the wall).  We will consider trajectories in which the detector travels from one wall to  the other  as well as from the midpoint as appropriate; we shall refer to the latter kind of trajectory as a `midpoint trajectory'.
    
    If the initial state of the field is the Minkowski vacuum state $\ket{0_M}$, then it is straightforward to show that the pullback of the Wightman function along the trajectory $\gamma(\tau)$ is 
    \begin{align}
        W_{0}(\tau,\tau')=\sum_{n=1}^\infty u_n(\sx(\tau))u_n^*(\sx(\tau'))\,.
    \end{align}
    For a uniformly accelerating detector, this trajectory is given by
    \begin{align}
        \sx(\tau) = \frac{1}{a}\rr{\sinh a\tau,\cosh a\tau -1}
    \end{align}
    where the integration constant is chosen so that $x(\gamma(0)) = 0$. Solving for the time taken to traverse the cavity, we obtain
    \begin{align}\label{tau-trav}
        \tau_{\text{max}} = \frac{\cosh^{-1}(1+aL)}{a}\,.
    \end{align}
    If the detector starts from the midpoint of the cavity, then the expression for the time to 
    exit the cavity is given by  Eq.~\eqref{tau-trav}  with  $L\to L/2$. Finally, putting everything together, we obtain 
    \begin{widetext}
    \begin{align}
        P^D_0(\Omega) = \lambda^2\sum_{n=1}^\infty\frac{1}{L\omega_n}\abs{\int_{-\infty}^{\infty} \dd\tau \chi(\tau)\sin\frac{n\pi}{L}\rr{\frac{\cosh a\tau-1}{a}-x_1}e^{-i\Omega \tau}\exp\rr{-i\omega_n\frac{\sinh a\tau}{a}}}^2 
    \end{align}
    \end{widetext}
 for the detector transition probability for the field in the vacuum state. Note that the limits of integration are
 effectively governed by the switching function.  We shall generally choose $\chi(\tau) = 1$ for the interval $[0,\tau_{\text{max}}]$ and zero otherwise (the so-called  top-hat switching).  We use the superscript $D$ to denote an accelerating detector in a cavity that is static with respect to the lab frame $(t,x)$; otherwise we use  a superscript $C$. Note that the cavity forces the field to be compactly supported in the interval $[x_1,x_2]$, beyond which the detector experiences no interaction with the field.
 
 We  remark that for a trajectory where the detector traverses the entire cavity (from one wall to another), the divergences associated with sudden switching do not occur because the Dirichlet boundary condition causes the field to vanish there (see for instance \cite{Lopp:2018cavity}). Effectively, the detector does not see the discontinuity in the switching. Furthermore, while divergences due to sudden switching arise in quite general contexts \cite{padmanabhan1996finite}, it is also now known that the spurious divergence due to sudden switching in Minkowski space is in fact dimension-dependent \cite{hodgkinson2012response} and the setup $(1+1)$D does not suffer this problem due to logarithmic nature of the singularity in the Wightman function. Since the mode sum is convergent even without a UV regulator, imposing UV cutoff is a computational convenience (cf. Appendix~\ref{appendix:convergence}). An IR cutoff naturally arises from the Dirichlet boundary condition; thus the usual divergence of a massless scalar field in (1+1) dimensions does not appear either.

	\section{Accelerating cavity}\label{sec:cavity}

	Now suppose we consider a rigid cavity of length $L$ as measured in the lab frame at $t=0$. The cavity is uniformly accelerating in the positive $x$-direction, and there is an inertial UDW detector at rest at $(t,x_d)$ where $x_d$ is constant. This corresponds to the detector passing through a cavity with moving boundary conditions. In (1+1) dimensions, there is an analytic solution to this seemingly difficult problem $m=0$: we perform a coordinate transformation
	\begin{align}
	    t = \frac{e^{a\zeta}}{a}\sinh a\varsigma\,,\hspace{0.25cm} x = \frac{e^{a\zeta}}{a}\cosh a\varsigma
	\end{align}
    where $(\varsigma,\zeta)$ are sometimes known as the Lass or radar Rindler coordinates \cite{lass:1963}  --- we will refer to these as \textit{conformal Rindler coordinates}. This coordinate system covers the usual Rindler wedge and has the special property that the metric is conformal to the Minkowski metric:
	\begin{align}
	    \dd s^2 = \dd t^2-\dd x^2 = e^{2a\zeta}\rr{\dd\varsigma^2-\dd\zeta^2}\,.
	\end{align}
	Each line of constant $\zeta$ describes a uniformly accelerating trajectory with proper acceleration 
    $|a^\mu a_\mu|^{1/2}=ae^{-a\zeta}$. Consequently the kinematical parameter $a$ for the line $\zeta=0$   corresponds to the proper acceleration of the test particle along this trajectory.	In these coordinates, the cavity walls correspond to Dirichlet boundary conditions at $\zeta=\zeta_1,\zeta_2$. Since we are comparing the scenarios in which the detector traverses the entire cavity, we will also choose $\zeta_1=0$ so that the proper acceleration of the left wall matches the acceleration of the detector\footnote{Note that if we consider the midpoint of the cavity to have acceleration $a$, then it is \textit{not} true that the walls are located at $\zeta_j=\pm L'/2$: conformal transformations do not preserve distances between two points. In particular, it can be shown that
	\begin{equation*}
	    x_j = \frac{1}{a}\pm \frac{L}{2} \Longrightarrow \zeta_j = \log\rr{1\pm \frac{aL}{2}}
	\end{equation*}
	which is manifestly not symmetric with respect to the detector position $\zeta_d=0$.}. Inverting the coordinates, the  trajectory of the static detector is
    \begin{align}
        \varsigma=\frac{1}{a}\tanh^{-1}\frac{t}{x_d}\,,\hspace{0.25cm}\zeta = \frac{1}{a}\log a\sqrt{x_d^2-t^2}\,.
    \end{align}
    If we define the left wall to be at $\zeta=\zeta_1=0$, then the proper length of the cavity in conformal coordinates is
    \begin{align}
    \label{lengthcavity}
        L = x_2-x_1\Bigr|_{t=0} = \int_{\zeta_1=0}^{\zeta_2:=L'} \!\!\dd\zeta\,e^{a\zeta} = \frac{e^{aL'}-1}{a}\,,
    \end{align}
    \vspace{0.15cm}
which can be inverted to give $L'=a^{-1}\log(1+aL)$. Crucially, $x_2-x_1\neq \zeta_2-\zeta_1$. If the detector starts at the right wall and the cavity accelerates in the positive $x$-direction, then we have $x_d=a^{-1}+L$. The maximum interaction time is obtained by solving for
    \begin{equation}
        \begin{aligned}
        a\sqrt{x_d^2-t^2} = 1 \Longrightarrow t_{\text{max}}=\sqrt{\frac{2L}{a}+L^2}\,. 
        \end{aligned}
    \end{equation}	
	For the massless field, the Klein-Gordon equation is conformally invariant under the above transformation and hence the modes in this coordinate system read
	\begin{align}
        \hat\phi(\varsigma,\zeta) &= \sum_{n=1}^\infty v_n(\zeta)\rr{e^{-i\tilde\omega_n \varsigma}\hat b_n+e^{i\tilde\omega_n \varsigma}\hat b_n^\dagger}\,,\\
        v_n &= \frac{1}{\sqrt{n\pi}}\sin\frac{n\pi(\zeta-\zeta_1)}{L'}\,,
    \end{align}
    where we have used the fact that the normalization simplifies due to $\sqrt{L'\tilde\omega_n} = \sqrt{n\pi}$.  
    Note that $L'\neq L$ since conformal transformation does not preserve length, i.e. the comoving length of the cavity in radar coordinates is $\zeta_2-\zeta_1\neq L$.

      
    Since $t=\tau$ is the proper time, the full transition probability for traversing the entire cavity is 
    \begin{widetext}
    \begin{align}\label{Prob-C}
        P^C_0(\Omega) = \lambda^2\sum_{n=1}^\infty\abs{\int_{-\infty}^{\infty} \dd\tau \chi(\tau)\sin\frac{n\pi\log\sqrt{(1+aL)^2-a^2\tau^2}}{\log{1+aL}}e^{-i\Omega \tau}\exp\rr{-\frac{in\pi\tanh^{-1}\frac{a\tau}{1+aL}}{\log(1+aL)}}}^2\,,
    \end{align}
    \end{widetext}
with the top-hat switching in the interval $[0,\tau_{\text{max}}]$, noting that here $t_{\text{max}}=\tau_{\text{max}}$.

 If the detector trajectory is such that at $t=0$ it is at midpoint of the cavity (`midpoint detector'), then some parts of these expressions will need to be changed if we want the kinematical parameter $a$ to be the proper acceleration at the centre of the cavity (such as is done in \cite{Dragan:2011zz}). Both $t_{\text{max}}$ and $L'$ will change for the midpoint detector
    \begin{equation}
        \begin{aligned}
        L' = \log\frac{2+aL}{2-aL}\,,\,\, t_{\text{max}} = \sqrt{\frac{L}{a}-\frac{L^2}{4}} 
    \end{aligned}
    \end{equation}    
    and there will be a  slight modification  of Eq.~\eqref{Prob-C}. Also, clearly $\zeta_1$ would not be zero in this case.   
    
    If the field is massive, the Klein-Gordon equation is no longer invariant under a conformal transformation, and it is more advantageous to use the manifestly simpler standard Rindler coordinates  
        \begin{equation}
        t = \xi\sinh \eta\,,\hspace{0.5cm} x= \xi\cosh\eta\,.
    \end{equation}
    Let us work this out explicitly from the Klein-Gordon equation: since $\sqrt{-g}=\xi$, the covariant Klein-Gordon equation gives
    \begin{equation}
        \frac{1}{\xi^2}\frac{\pd^2\phi}{\pd\eta^2} - \rr{\frac{1}{\xi}\frac{\pd\phi}{\pd\xi}+\frac{\pd^2\phi}{\pd\xi^2}}+m^2\phi=0\,.
    \end{equation}
Separating  variables $\phi=v(\xi)T(\eta)$, we can show that $T(\eta)\propto \exp \pm i\omega\eta$ and hence we obtain the modified Bessel differential equation of imaginary order for the spatial mode $v(\xi)$:
    \begin{equation}
        \xi^2\frac{\pd^2v}{\pd\xi^2}+ \xi\frac{\pd v}{\pd\xi}+ (\omega^2-m^2\xi^2)v=0\,.
    \end{equation}
    Implementing the Dirichlet boundary condition $v(\xi_1)=v(\xi_2)=0$ as before, the modes will have discrete spectrum labelled by $n\in \Z$ and the spatial mode can be expressed
    in terms of modified Bessel functions of imaginary order \cite{NIST:DLMF}:
    \begin{equation}
        \begin{aligned}
        v_n(\xi) &= |A_n|\left(\text{Re}\rr{I_{i\omega_n}(m\xi_1)}K_{i\omega_n}(m\xi) - \right.\\
        &\hspace{0.5cm}\left.\text{Re}\rr{I_{i\omega_n}(m\xi)}K_{i\omega_n}(m\xi_1)\right)\,,\\
        1 &= 2|A_n|^2\omega_n\int_{\xi_1}^{\xi_2} \frac{\dd\xi}{\xi} \abs{v_n(\xi)}^2\,.
        \label{Besselbasis}
        \end{aligned}
    \end{equation}
    where the normalization follows from Klein-Gordon inner product in Eq.~\eqref{KGinnerproduct}. The discrete spectrum and the normalization must be solved numerically. Similar to the massless case, we can then do the pullback of the Wightman function onto the trajectory of the detector which is given by
    \begin{equation}
    \begin{aligned}
        \xi(\tau) &= \sqrt{x_d^2-t^2} \qquad 
        \eta(\tau) = \tanh^{-1}\frac{t}{x_d} 
    \end{aligned}
    \end{equation}
 where the constant $x_d$ describes  the static detector trajectory with respect to the lab coordinates.
    
	We pause to comment about  rigid body motion in the Rindler frame. Note that even if the leftmost wall acceleration gets arbitrarily large, the centre of mass acceleration is bounded above by the rigidity condition:  a rigid cavity of length $L$ in the lab frame  must have a different proper acceleration at each point in order to remain rigid.
The proper acceleration at any point $x$ within the cavity 
is given by
	\begin{align}
	    a(x) = \frac{a_1}{1+a_1(x-x_1)}\,,
	\end{align}
	where $x_1=a_1^{-1}$ is the location of left wall and $a_1$ is the proper acceleration of the left wall. We see that at the centre  $x_c = a_1^{-1}+L/2$ of the cavity we have the limit
		\begin{align}
	    \lim_{a_1\to\infty} a_c = \lim_{a_1\to\infty}\frac{2a_1}{2+a_1L} = \frac{2}{L}\,.
	\end{align}
If the centre of the cavity attains an acceleration larger than this, the rear wall will cross the future Rindler horizon, which is an unphysical cavity setup.

 Another way to see this geometrically is by looking at the spacetime diagram (cf. Figure~\ref{rindlerdiagram}). 
    For a uniformly accelerating rigid cavity, the two walls must both be on two different hypersurfaces of constant $\xi$ in order for them to be a Dirichlet boundary i.e. $\xi=\xi_1$ and $\xi=\xi_2$. Different values of $\xi$ correspond to trajectories with different proper accelerations, and the lab observer does not see this cavity as rigid because the the cavity shrinks across plane of simultaneity of constant $t$. The rigidity condition essentially means that cavity has constant length when measured in the plane of simultaneity of constant $\eta$.

\begin{figure}[htp]
	    \centering
	    \includegraphics[scale=0.9]{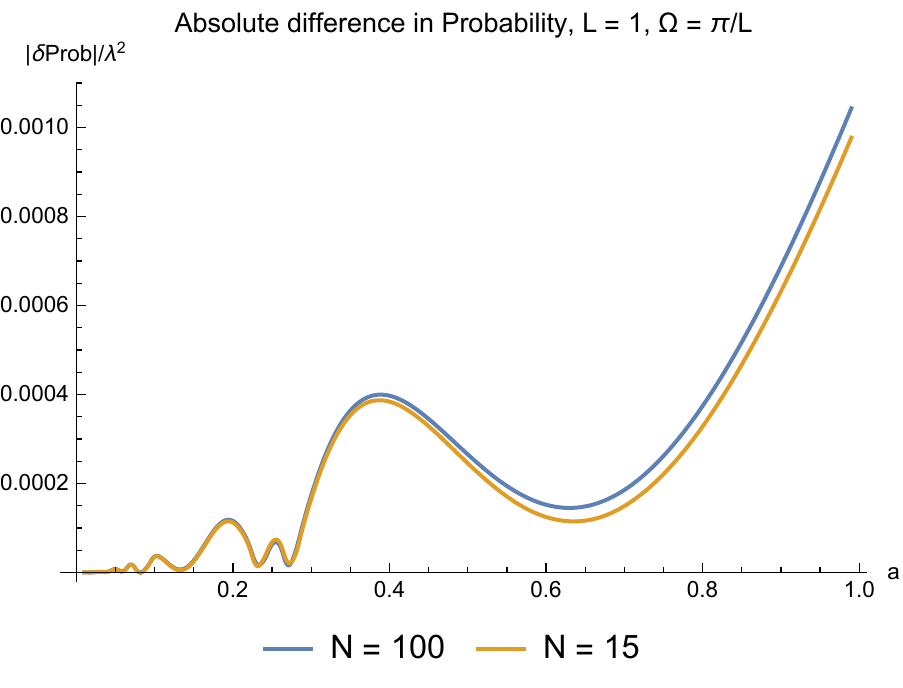}
	    \caption{Absolute difference in probability $|P^C_0-P^D_0|/\lambda^2$ as a function of acceleration for $\Omega = \pi/L$ for $M=0$. Here and for subsequent plots we set $L=1$ for convenience. For small accelerations, the mode sums quickly converge for small $N$ and the difference in transition probability of the two scenarios is vanishingly small in low acceleration limit.}
	    \label{plotdiff1}
	\end{figure}

\begin{figure}[htp]
	    \centering
	    \includegraphics[scale=0.9]{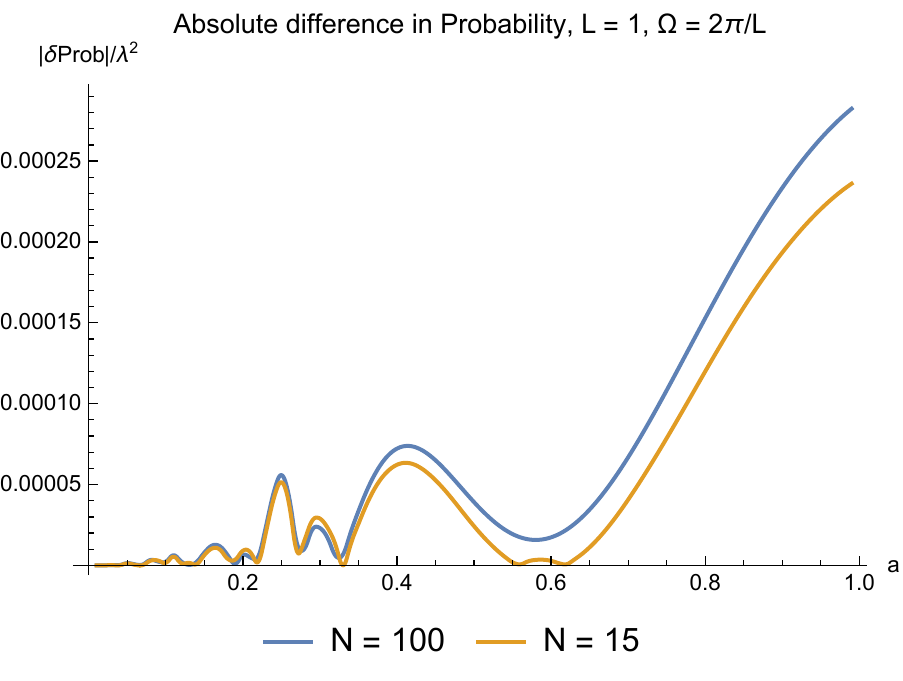}
	    \caption{Absolute difference in probability $|P^C_0-P^D_0|/\lambda^2$ as a function of acceleration for larger gap $\Omega = 2\pi/L, L = 1$ for $m=0$.}
	    \label{plotdiff2}
	\end{figure}

	In Figure~\ref{plotdiff1} we plot the absolute probability difference between the accelerating cavity and the accelerating detector scenarios as a function of the proper acceleration $a$. A larger energy gap generally suppresses the transition probability in massless scenario  as shown in Figure~\ref{plotdiff2}, and similar qualitative suppression is observed in the massive case.
    For comparison of the convergence of the  mode sums, we considered ranging both $N=15$ and $N=100$. The larger value of $N$ is required for larger acceleration parameters $a$ (see also Appendix~\ref{appendix:convergence} for separate convergence checks). 

\begin{figure}[htp]
	    \centering
	    \includegraphics[scale=0.8]{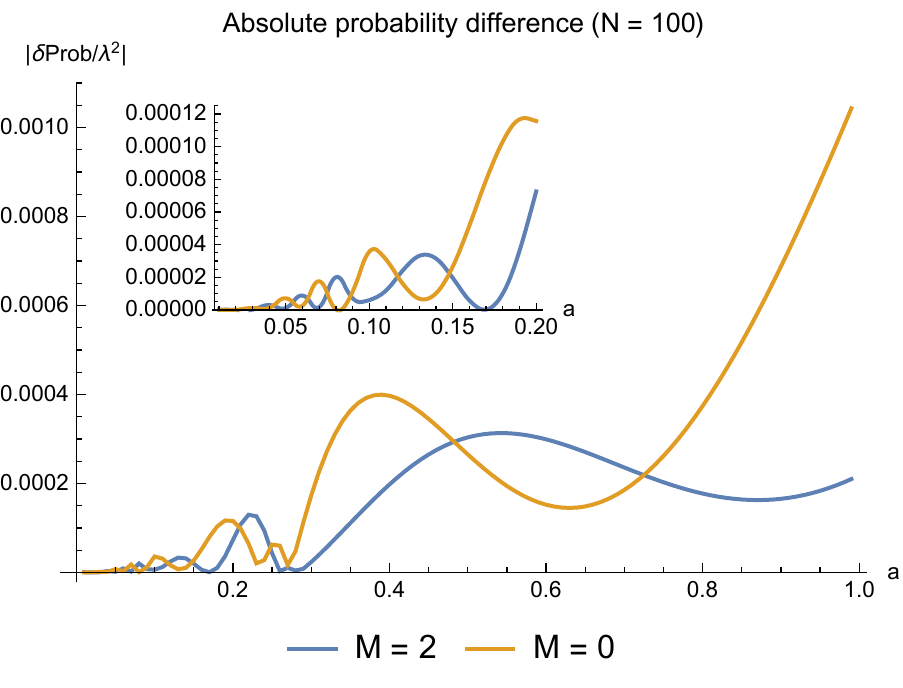}
	    \caption{Comparing the absolute difference in probability $|P^C_0-P^D_0|/\lambda^2$ as a function of acceleration for $L=1,\Omega = \pi/L$ when the field is initially in the vacuum state.  Here $L = 1$ for convenience. The difference between an accelerating cavity and an accelerating detector vanishes quickly at low accelerations. 	    }
	    \label{plotdiff3}
	\end{figure}

\begin{figure}[htp]
	    \centering
	    \includegraphics[scale=0.9]{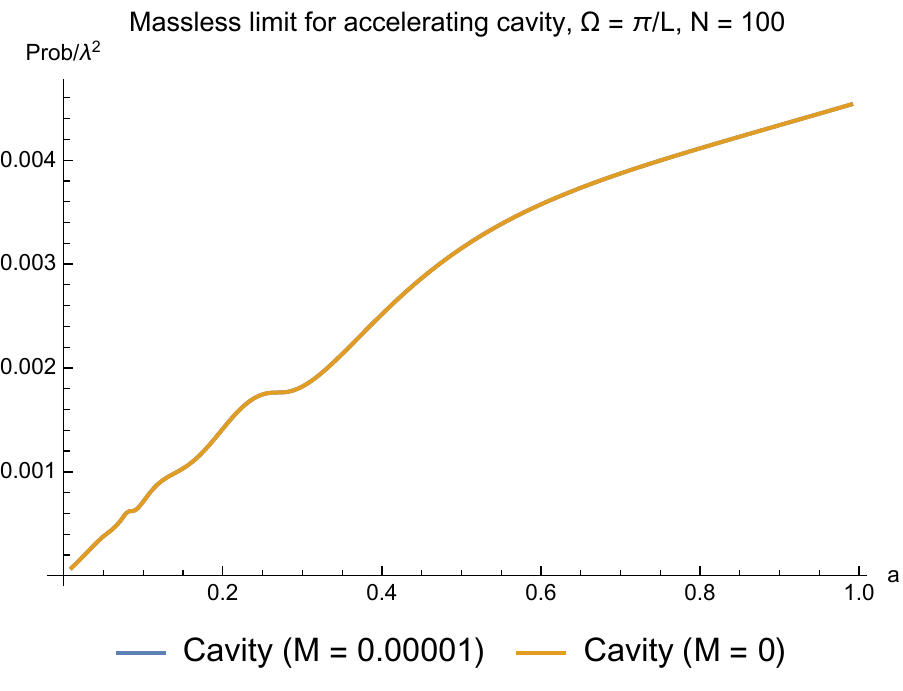}
	    \caption{Transition probability (divided by $\lambda^2$) as a function of acceleration for $\Omega = \pi/L, L = 1$, showing that in the small mass limit the results agree with massless case. We choose $N=100$  instead of the value $N=15$ as in previous plots. Note that a value of $M=0.0001$ is small enough to be indistinguishable from the
	    $M=0$ case,  with relative difference in probability of 2 parts in a billion ($10^{-9}$) at $a\approx 0.01$.
	    }
	    \label{masslesslimit}
	\end{figure}

	In Figure~\ref{plotdiff3} we compare the absolute probability difference for massless and massive fields.  Here our results agree with previous work \cite{Dragan:2011zz} in that if the initial field state is the vacuum, then for $aL\ll 1$ the difference in responses between inertial and non-inertial detectors quickly vanishes. For completeness, we plot in Figure~\ref{masslesslimit} the transition probability of the accelerating cavity scenario for very small mass.  We see that indeed it provides the correct massless limit despite the rather complicated mode functions involving modified Bessel functions. The accelerating detector case will trivially have the correct limit since the functional form of the Wightman functions is the same.
	
    Furthermore, we do see considerable distinction between the massive and massless cases once $a$ becomes sufficiently large (cf. Figure~\ref{plotdiff3}). The larger difference in response for the two setups at large $a$ should not be taken to be a fundamental violation of the WEP: for large $a$, the non-uniformity of the cavity acceleration at different points  is more pronounced, similar to how the non-uniformity of Earth's gravitational field  is detectable if we consider a large enough region in space.
    
	\section{Excited field states}\label{sec:excited}
	
	After considering the vacuum state of the field, a natural question then arises: can sensitivity to non-inertiality be enhanced if the field state is not a vacuum state? The additional terms in the Wightman function due to the excited field states may have co-rotating term of the form $\Omega-\omega_n$ which may produce resonant-like behaviour, while for the vacuum state this cannot occur for a ground state atom. We will consider both single-mode excited Fock states and single-mode coherent states.
	
	\subsection{Single-mode excited Fock state}
	The simplest excited field state we can consider is a single-mode non-vacuum Fock state, i.e. when the $k$-th momentum has $n_k$ excitations. This is a straightforward generalization from the expression found in \cite{Dragan:2011zz}. We denote this by $\ket{n_k}$ which formally reads $\ket{n_k}\sim\ket{000... 0\;n_k\;000...}$,
	where the enumeration is formally valid because of the countably infinite spectrum. The corresponding Wightman function is formally given by
	\begin{equation}
	    \begin{aligned}
	    W(\sx,\sx') &= \braket{n_k|\phi(\sx)\phi(\sx')|n_k} \\
	    \end{aligned}
	\end{equation}
Employing the result
	\begin{equation}
	    \begin{aligned}
	    \phi(\sx')\ket{n_k}
	    &=\sum_{l\neq k}u_l^*(\sx')\ket{1_l,n_k} +\\
	    &\hspace{0.5cm}\sqrt{n_k+1}u_l^*(\sx')\ket{n_k+1}+ \\
	    &\hspace{0.5cm}\sqrt{n_k}u_l(\sx')\ket{n_k-1} 
	    \end{aligned}
	\end{equation}
    where the $\{u_j\}$ are the eigenmodes of the Klein-Gordon equation (not just the spatial part), we obtain 
    \begin{equation}
	    \begin{aligned}
	        W(\sx,\sx') &= \sum_{j,l\neq k}
	        u_j(\sx)u_l^*(\sx')+ (n_k+1)u_k(\sx)u_k^*(\sx')\\
	        &\hspace{2.95cm}  +n_k u^*_k(\sx) u_k(\sx')\\
	        &= \sum_j u_j(\sx) u_j^*(\sx') + n_k u_k(\sx) u_k^*(\sx')\\
	        &\hspace{2.8cm}  + n_ku_k^*(\sx) u_k(\sx')\\
	        &= W_0(\sx,\sx') + W_{\text{exc}}(\sx,\sx')
	    \end{aligned}
	\end{equation}
for the full expression for the Wightman function.
		Therefore, for an excited field state given by a single-mode Fock state, the Wightman function is the sum of the vacuum Wightman function $W_0$ and an additional piece $W_{\text{exc}}$ that is explicitly dependent on which mode it is excited. Since the transition probability is linear in $W(\sx,\sx')$, we see that the transition probability for this state reads
	\begin{equation}
	    \begin{aligned}
	    P_{\text{tot}}(\Omega) &= P_0(\Omega) + n_k\abs{\int\dd\tau\chi(\tau)e^{-i\Omega\tau}u_k(\tau)}^2 + \\ &\hspace{1.8cm} n_k\abs{\int\dd\tau\chi(\tau)e^{-i\Omega\tau}u_k^*(\tau)}^2 \,.
	    \label{resonantprob}
	    \end{aligned}
	\end{equation}
    We are interested in $W_{\text{exc}}$ since we found $W_0$ in the previous section and we can always subtract off the vacuum contribution. Note that
    \begin{equation}
    \begin{aligned}
        e^{-i\Omega\tau}u_k(\tau) &\sim e^{-i(\omega_k T(\tau)+\Omega\tau)}\,,\\
        e^{-i\Omega\tau}u_k^*(\tau) &\sim e^{-i(\omega_k T(\tau)-\Omega\tau)}\,,
    \end{aligned}
    \end{equation}
    where $T(\tau)$ is the time function (which in our case is either $\eta(\tau)$ or $t(\tau)$) along the trajectory of the detector. The third term in $P_{\text{tot}}$ is the `co-rotating term' which will tend to dominate over the second (`counter-rotating') term.
    
    The above results teach us that there are two ways in which one can ``neglect" the vacuum contribution.  One is when we have an approximate \textit{resonance} (up to some Doppler shifts) i.e. when $\Omega \sim \omega_k $. In this case, the resonance will amplify the transition rate and the vacuum contribution can be rendered  negligible compared to the rest. The other is if there is a sufficiently higher number of excitations $n_k$: in this case the transition probability scales as
    \begin{align}
        P_{\text{excited}} \sim \frac{n_k}{k}
    \end{align}
    where the denominator $1/k$ comes from the normalization of $u_k$. This means for a given energy gap $\Omega$, the higher momentum mode will need an excitation of order $n_k\sim k$ to achieve a given probability amplitude. When it is off-resonance, a larger gap tends to diminish the transition probability, which simply reflects the fact that atoms with larger energy gaps are harder to excite.
    
\begin{figure}[htp]
        \centering
        \includegraphics[scale=0.9]{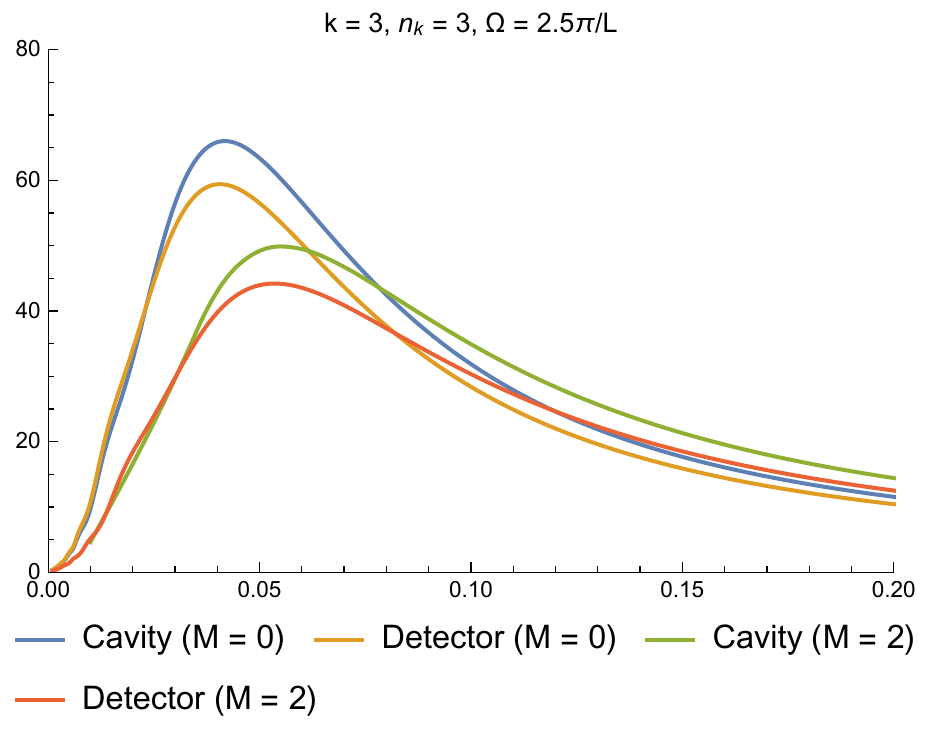}
        \includegraphics[scale=0.9]{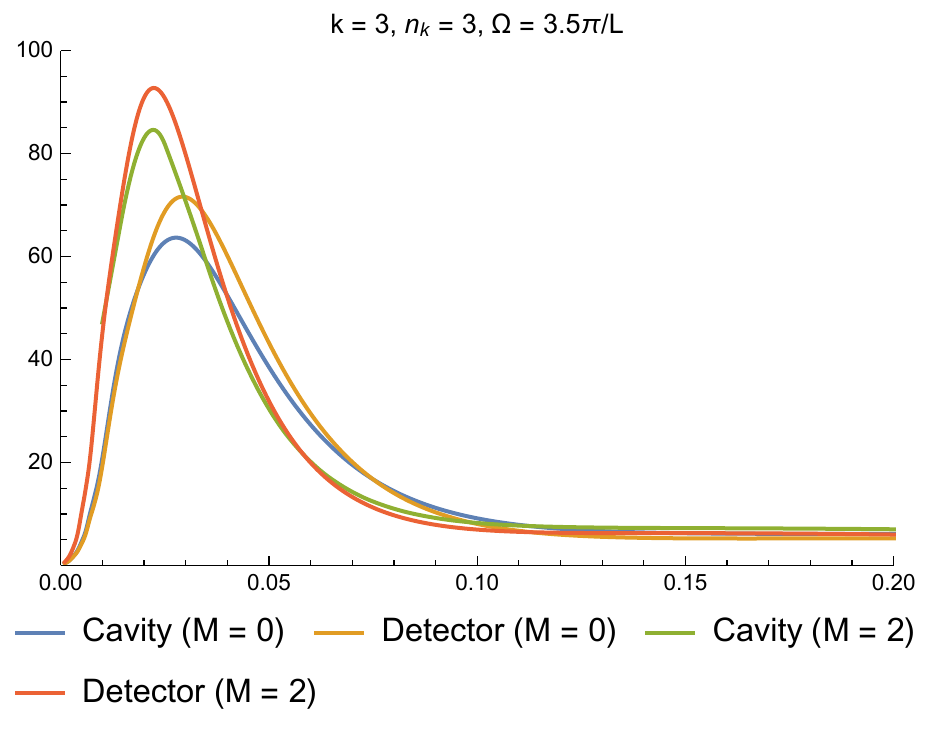}
        \caption{Transition probability (divided by $\lambda^2$) as a function of acceleration for two different gaps, comparing massless and massive cases. The field is in the third excited state i.e. $k=3$ and we chose $n_3 = 3$. \textbf{Top:} $\Omega = 2.5\pi/L$. \textbf{Bottom:} $\Omega=3.5\pi/L$. The plots are for $L = 1$. }
        \label{plotexcited1}
    \end{figure}

    Some of these results are shown in Figure~\ref{plotexcited1}. A notable result upon comparison of the two figures is that one can indeed amplify transition probability by considering gaps that are `close' to the excited field state frequency. 
    In Figure~\ref{plotexcited1}, by considering `off-resonant' gap at $\Omega=3\pi/L\pm \epsilon$,  there are regimes of accelerations in which the massive fields have better transition probabilities for both accelerating detector/cavity scenarios than do their massless counterparts, and vice versa depending whether $\Omega=\omega_n-\epsilon$ or $\Omega=\omega_n+\epsilon$ (in the plots, $\epsilon=0.5\pi/L)$. However, for each mass the distinction between an accelerating detector and an accelerating cavity quickly vanishes for small $a$. 
    
    Here we make a parenthetical comment that the relative magnitude of $\Omega-\omega_k$ or $\Omega/\omega_k$ \textit{does matter}: for a given fixed atomic gap $\Omega$, one can engineer a situation in which massive fields can have larger transition probability than the massless counterpart using resonance and vice versa. This is already apparent in Figure~\ref{plotexcited1} for small $a$, where transition probability for massive case can be lower or higher than the massless case depending on choice of gap $\Omega$. This is, however, a separate problem from fundamentally distinguishing local accelerations.
    
\begin{figure}[htp]
        \centering
        \includegraphics[scale=0.9]{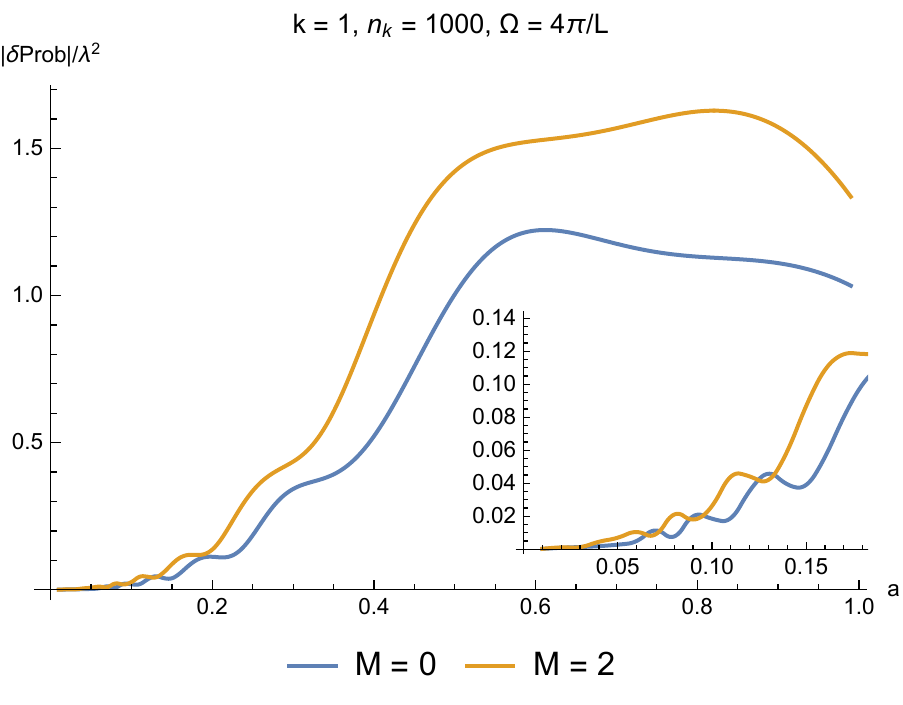}
        \includegraphics[scale=0.9]{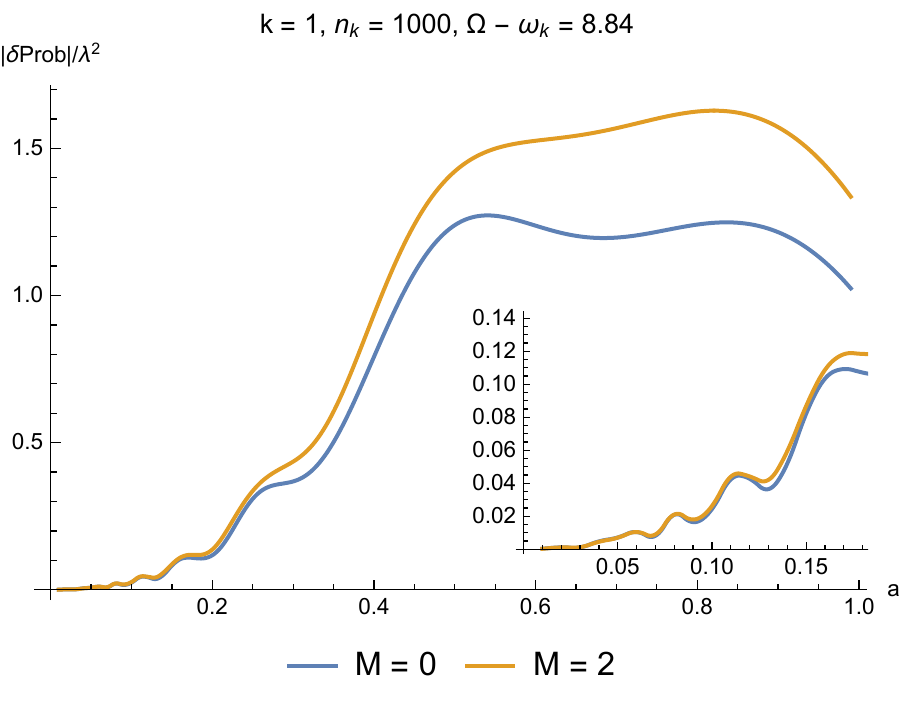}
        \includegraphics[scale=0.9]{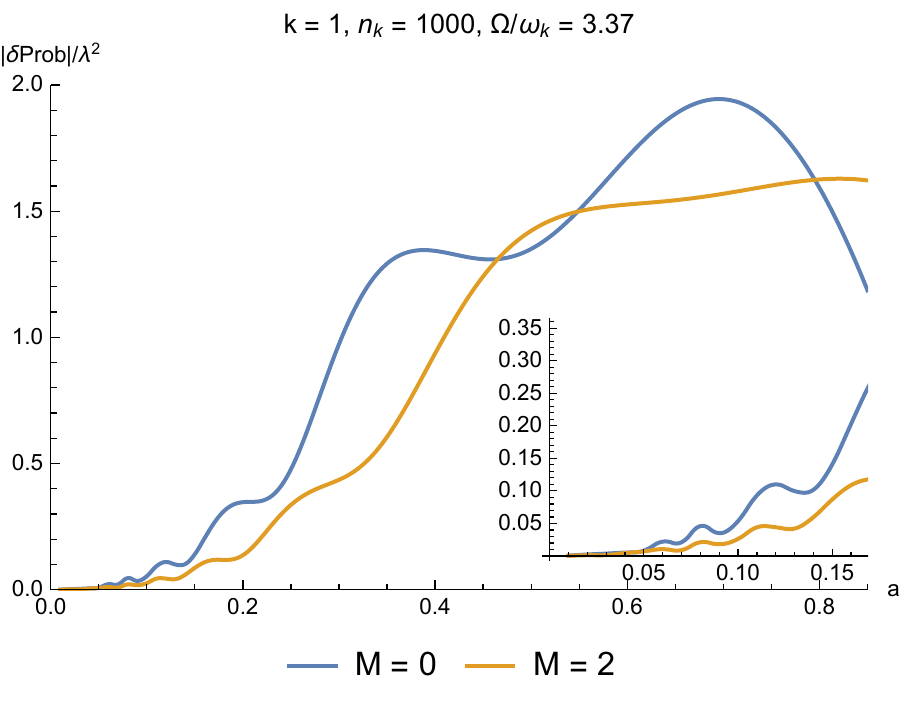}
        \caption{Absolute probability difference (divided by $\lambda^2$) as a function of acceleration for large $n_k$. The field is in the third excited state i.e. $k=3$ and we chose $n_3 = 3$. \textbf{Top:} $\Omega = 4\pi/L$. \textbf{Middle:} $\Omega-\omega_k = 8.84$ where the reference $\omega_k$ is chosen to be the angular frequency for the massive case. \textbf{Bottom:} $\Omega/\omega_k=3.37$. The plots are for $L = 1$. }
        \label{plotexcited2}
    \end{figure}

    The relative magnitude matters less as one moves away from resonance, e.g. when $\Omega/\omega_k\gg 1$. We check this for the case of highly populated field state $n_k\gg 1$ as shown in Figure~\ref{plotexcited2}, where we choose $k = 1$ and $n_k = 1000$ to match the setup in \cite{Dragan:2011zz} for convenience. In the top figure, the massive field seems to outperform the massless case for distinguishing local accelerations. However, this can be attributed to resonant effect, since for our choice of fixed $\Omega$, the magnitude of $\Omega-\omega_k$ is smaller for the massive case than the massless case. A possibly fairer comparison would be to use the same $|\Omega-\omega_k|$ or $\Omega/\omega_k$, as shown in the middle and bottom plot of Figure~\ref{plotexcited2}. When this is done, we see that the apparent advantage of the massive field over massless one disappears and massless field seems to perform equally well if not better\footnote{This issue is somewhat tricky since it may arguable which comparison is fairer. However, this `fairness' is necessary for WEP since fair comparison is analogous to ``not being able to look out of the window of a rocket" to decide the asymmetry of the problem.}.
    
    We conclude from these that massive fields do not seem to offer any obvious  fundamental  advantages at low accelerations as compared to their massless counterpart. 
    In Appendix~\ref{appendix:disparity} we suggest a possible reason for the disparity with the results found in \cite{Dragan:2011zz}.

    \subsection{Coherent field state}
    
    An interesting case to consider is when the field is in a coherent state, analogous to that of a laser field in quantum optics scenarios. It is defined as the continuum limit of a quantum-mechanical coherent state for a quantum harmonic oscillator using the displacement operator $\hat D_{\alpha(k)}$ (see, for instance \cite{Simidzija:2017jpo}):
    \begin{equation}
        \begin{aligned}
        \ket{\alpha(k)} &:= \hat D_{\alpha(k)}\ket 0 \\
        &= \exp\left[\int \dd k\com{\alpha(k) \hat a^\dagger_k - \alpha^*(k) \hat a_k}\right]\ket 0\,.
    \end{aligned}
    \end{equation}
  Here $\alpha(k)$ is the \textit{coherent amplitude distribution} defining a coherent amplitude for every mode $k$. As a coherent state, it satisfies the `eigenvalue' equation
    \begin{align}
    \label{coherentannihilate}
        \hat a_{k'}\ket{\alpha(k)} = \alpha(k')\ket{\alpha(k)}\,,
    \end{align}
    noting that $\ket{\alpha(k)}$ does not mean an explicit dependence on $k$ but rather on coherent amplitude distribution $\alpha(k)$.
    In a cavity, the spectrum becomes discrete and so we label the modes with integers $n$ instead (for example, the continuous variable $k$ becomes discrete: $k_n=n\pi/L$ in static cavity scenario). The coherent state  has a simpler form
    \begin{align}
        \ket{\alpha(n)} = \exp\left[{\sum_{n=1}^\infty (\alpha_n \hat a^\dagger_n - \alpha^*_n \hat a_n})\right]\ket 0\,.
    \end{align}
    Note that in this case we can formally write
    \begin{equation}
        \ket{\alpha(n)}\sim \ket{\alpha_1\alpha_2...\alpha_j...} \sim \bigotimes_{n=1}^\infty \ket{\alpha_n}
    \end{equation}
    which denotes tensor product of coherent states each with complex coherent amplitude $\alpha_j$. For single-mode coherent state, say for the $j$-th momentum, we have (cf. Eq.~\eqref{coherentannihilate})
    \begin{align}
        \hat a_j\ket{\alpha(k)} = \alpha_j\delta_{jk}\ket{\alpha(k)}\,,\hspace{0.25cm}\alpha_j\in\C\,.
    \end{align}
    For a countably infinite multimode coherent state above, we require that
    \begin{align}
        \sum_{n=1}^\infty \abs{\alpha_n}^2 < \infty
    \end{align}
    which means that modes with higher momenta have suppressed coherent amplitude.  Here we will not employ the infinite multimode coherent state,  and instead focus specifically on the more realistic single-mode coherent state as is used in quantum optics.
    
    The Wightman function for the coherent state reads
    \begin{equation}
        \begin{aligned}
        W(\sx,\sx') &= \braket{0|\hat D^\dagger_{\alpha(n)}\phi(\sx)\phi(\sx')\hat D_{\alpha(n)}|0} \\
        &= \sum_{n=1}^\infty u_n(\sx)u_n^*(\sx') + \sum_{n=1}^\infty\sum_{j=1}^\infty \alpha_n^*\alpha_j u_j(\sx)u_n^*(\sx')\\
        &\hspace{0.25cm} + \sum_{n=1}^\infty\sum_{j=1}^\infty \alpha_n\alpha_j^* u_j^*(\sx)u_n(\sx')\\
        &\hspace{0.25cm} + \sum_{n=1}^\infty\sum_{j=1}^\infty \alpha_n\alpha_j u_j(\sx)u_n(\sx')\\
        &\hspace{0.25cm} + \sum_{n=1}^\infty\sum_{j=1}^\infty \alpha_n^*\alpha_j^* u_j^*(\sx)u_n^*(\sx')\,.
        \end{aligned}
    \end{equation}
    Note that similar to the single-mode excited Fock state, the vacuum contribution to the Wightman function does not vanish. If we define the one-point function of the coherent state as
    \begin{align}
        J(\sx):=\braket{\alpha(n)|\phi(\sx)|\alpha(n)} = \sum_n \alpha_n u_n(\sx)\,,
    \end{align}
    we can compactly write the full Wightman function as
    \begin{equation}
        \begin{aligned}
        W(\sx,\sx') &= W_0(\sx,\sx') + J(\sx)J(\sx') + J(\sx)J^*(\sx') \\
        &\hspace{1.9cm} + J^*(\sx)J(\sx') + J^*(\sx)J^*(\sx')\\ 
        &= W_0(\sx,\sx')+W_{c}(\sx,\sx')\,,\\
        W_{c}(\sx,\sx') &= 4\text{Re}[J(\sx)]\text{Re}[J(\sx')]\,.
        \end{aligned}
    \end{equation}
    
    The fact that $W_c(\sx,\sx')$ factorizes into product of one-point functions allow us to simplify the expression for the transition probability. The transition probability due to the purely  coherent part (i.e. modulo the vacuum contribution $W_0(\sx,\sx')$) then reads
    \begin{equation}
        \begin{aligned}
        P_c(\Omega) &= \lambda^2\int\dd \tau\,\dd\tau'\chi(\tau)\chi(\tau')e^{-i\Omega(\tau-\tau')}W_c(\tau,\tau')\\
        &= 4\lambda^2\abs{\int\dd \tau \chi(\tau)e^{-i\Omega\tau}\text{Re}[J(\sx(\tau))]}^2\,.
    \end{aligned}
    \end{equation}
    With a judicious choice of $\{\alpha_n\}$, it may be possible to perform the infinite sum in $J(\tau)$ exactly. Before we proceed, it is worth noting that resonant behaviour similar to that of the previous section is expected, since the real part of $J(\tau)$ contains $\cos \omega_n t(\tau)$ term which produces co-rotating term when combined with the exponential of the gap $e^{-i\Omega \tau}$.
    
    For single-mode coherent state, there is no real restriction on the coherent amplitude; we obtain
    \begin{align}
        J(\sx) = \delta_{mn}\sum_n\alpha_nu_n(\sx) \,,
    \end{align}
    where $m$-th mode is to be the coherent state and the rest are all vacuum modes. For simplicity we can consider, for example, $m=2$ and restrict $\alpha\in \R$ (though $\alpha $ can be arbitrary complex number).
    
\begin{figure}[htp]
        \centering
        \includegraphics[scale=0.8]{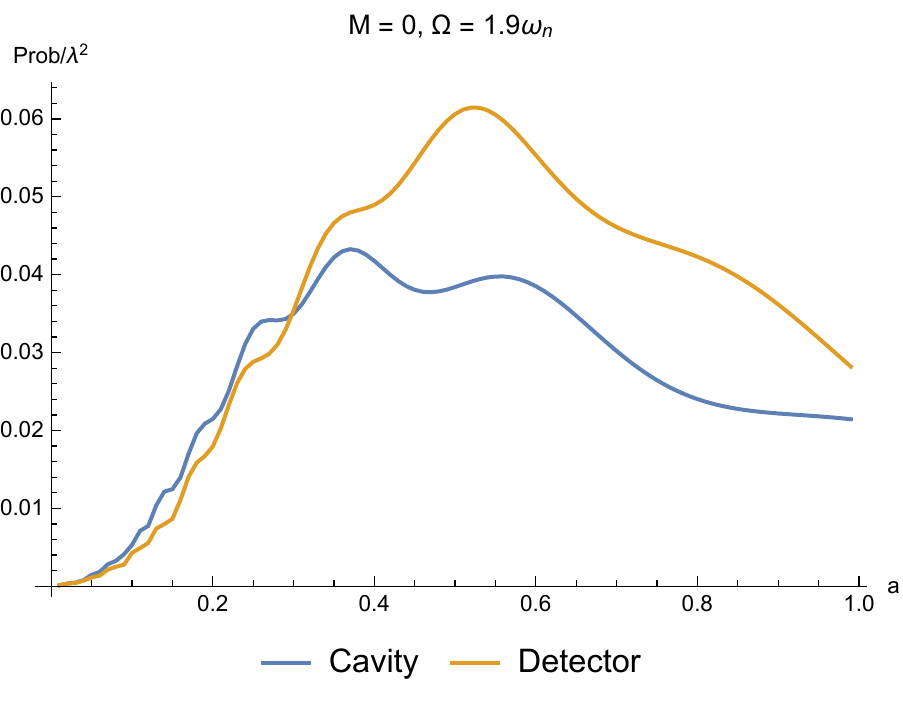}
        \includegraphics[scale=0.8]{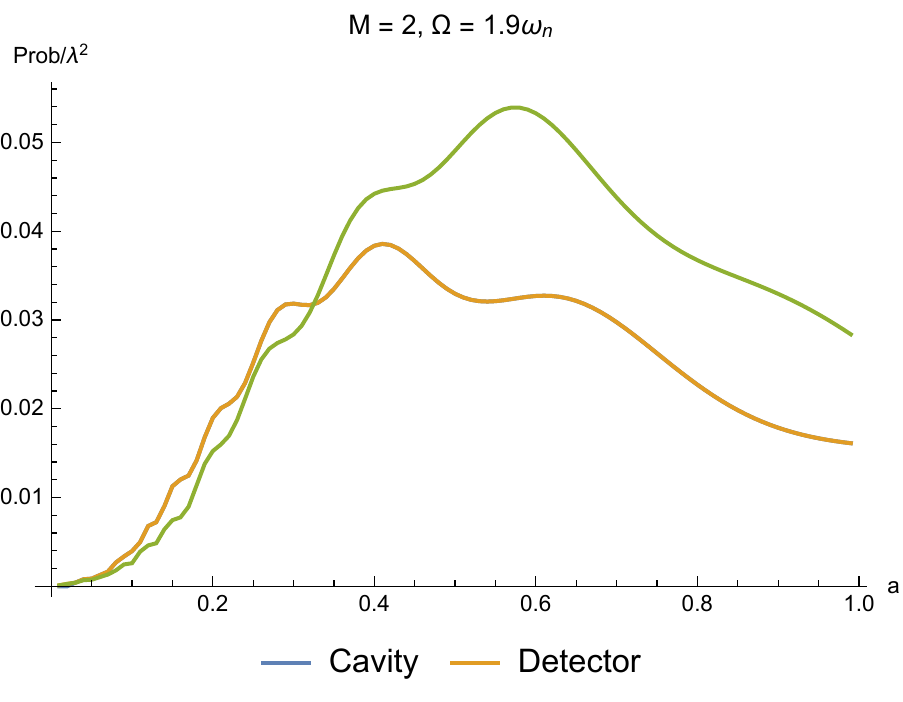}
        \caption{Transition probability  (modulo the vacuum contribution) for an accelerating detector and an accelerating cavity for two different masses when the second mode ($k=2$) is in coherent state and other modes are in vacuum state. In these plots $L = 1$.  }
        \label{plotsinglecoherent1}
    \end{figure}

\begin{figure}[htp]
        \centering
        \includegraphics[scale=0.8]{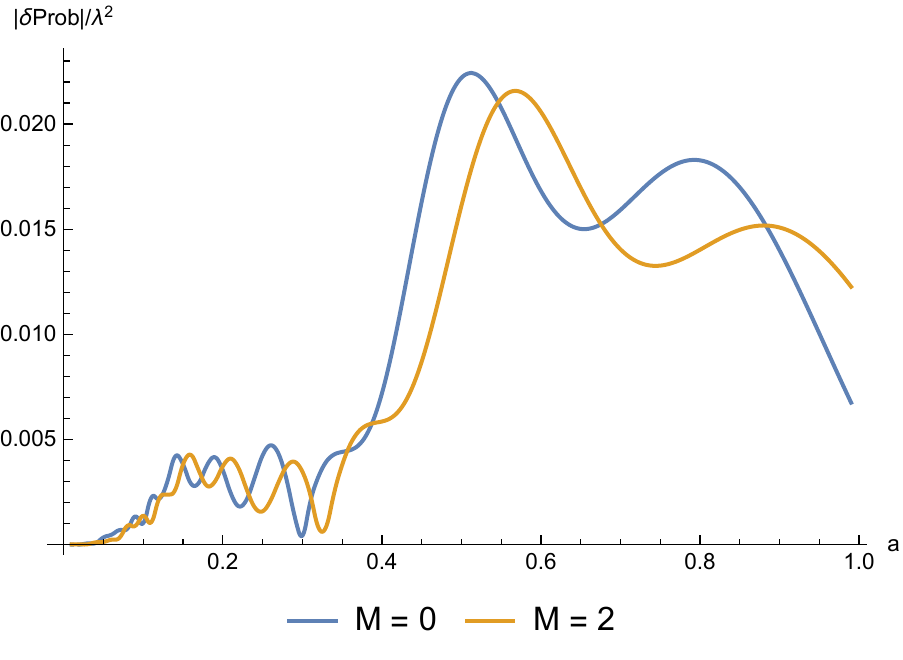}
        \includegraphics[scale=0.8]{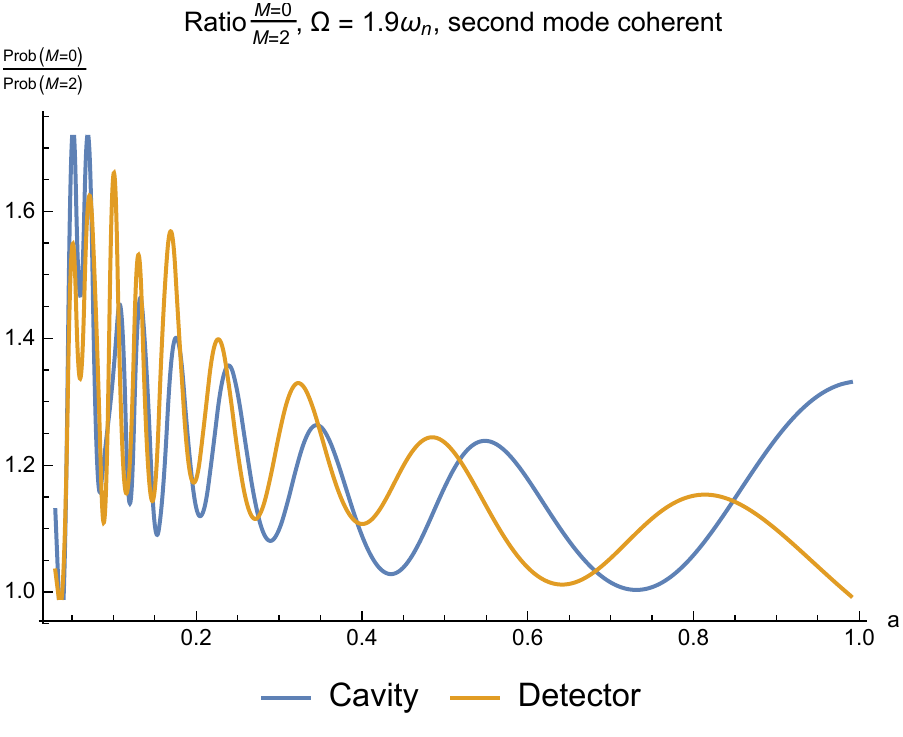}
        \caption{\textbf{Top:} absolute probability difference $|\delta P|/\lambda^2=|P^C(\Omega)-P^D(\Omega)|/\lambda^2$  (modulo the vacuum contribution) for massless and massive fields. The difference is vanishing for small $a$ regardless of mass. \textbf{Bottom}: the ratio of probabilities between massless and massive ones. We see that in low acceleration regimes the ratio approaches 1. }
        \label{plotsinglecoherent2}
    \end{figure}

    We illustrate the case when the second mode $k=2$ is in a coherent state with coherent amplitude $\alpha_2 = 1$ while others are in the vacuum state, shown in Figure~\ref{plotsinglecoherent1}. We also intentionally adjust the energy gap of the detector so that $\Omega=1.9\omega_n$, which is different for massless and massive fields. This comparison can be thought of as making the comparison somewhat fairer since the amount by which the atom is off-resonant from the mode frequency is of the same weight. We see that even with massive fields, the overall behaviour remains unchanged and as expected, the transition amplitude degrades with larger mass. This contrast is even more apparent when we compute absolute probability difference between the accelerating cavity and accelerating detector in massive and massless fields, as shown in Figure~\ref{plotsinglecoherent2}.  While we do not probe extremely non-relativistic regimes due to computational resources, it is clear that the role of mass is vanishingly small for smaller acceleration. We have ignored the vacuum contribution because we have chosen the value of $\Omega$ such that the vacuum contribution is negligible compared to the contribution due to the excited field state. Furthermore, we have shown that vacuum states are not sensitive to local accelerations.
 
    We pause  to comment that the response of an accelerating cavity response   `underperforms' relative to an accelerating detector for a fixed mass $m$ for large accelerations.  We see in  Figure~\ref{plotexcited1} that this under/overperformance is reversed for  $a\lesssim 0.35$.  This can presumably be  attributed  to non-linearities introduced by non-uniform acceleration across the accelerating cavity,  though we do not yet have a full understanding of this effect. 
    
    \subsection{Resonance}\label{sec:resonance}
    
\begin{figure}
        \centering
        \includegraphics[scale=0.9]{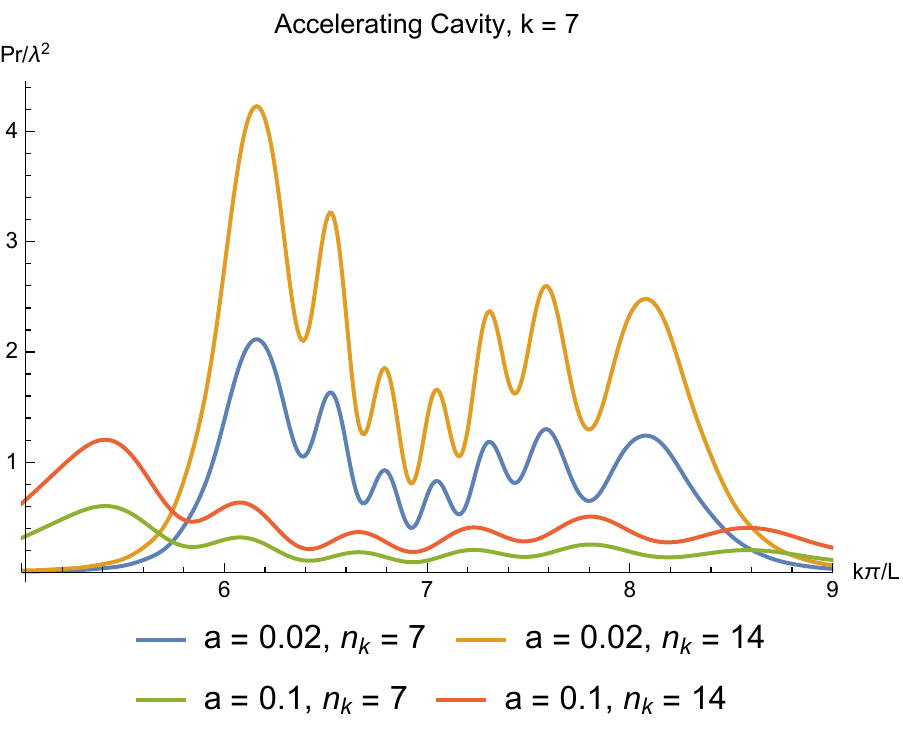}
        \includegraphics[scale=0.9]{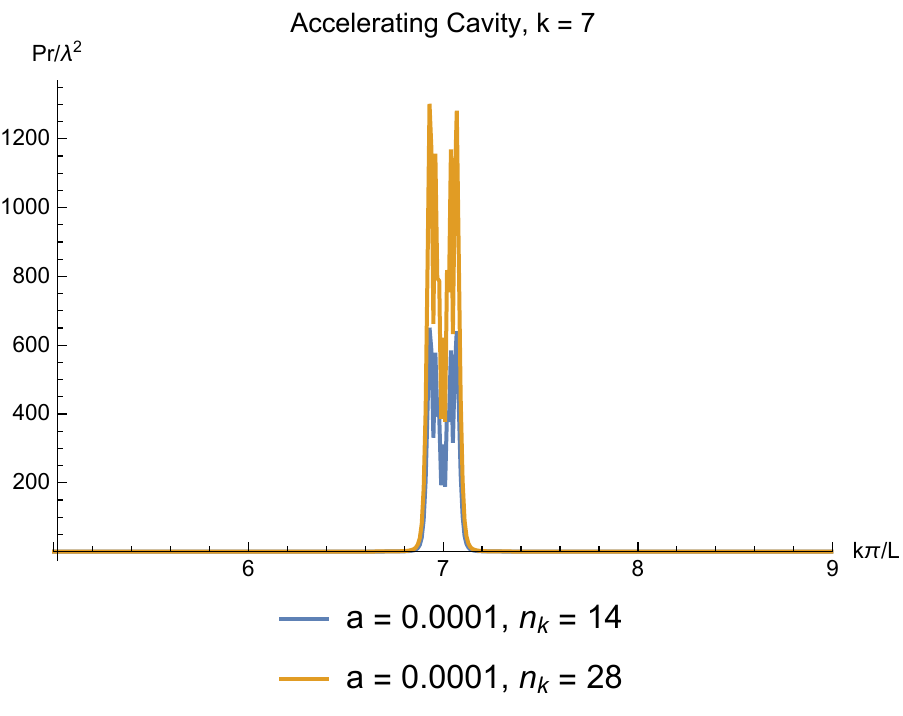}
        \caption{Transition probability for fixed acceleration as a function of energy gap $\Omega = k\pi/L$ where $n$ is real and $n_k$ is the number of excitations of the massless scalar field in mode $k$, with $L=1$.}
        \label{plotresonance1}
    \end{figure}

    The resonance phenomenon, while not very exact due to accelerated motion of the detector or cavity, can be made manifest if we study the ``resonance peak" of the detector. The resonance peak for the case of the field in a Fock state $\ket{n_k}$ is shown in Figure~\ref{plotresonance1}. Recall that in this notation, it is the $k$-th momentum having $n_k$ excitations: if field is in the seventh excited state with 20 excitations, then we write $\ket{20_7}$.
    
    From Figure~\ref{plotresonance1} we observe that for large acceleration there is a larger Doppler shift, which smears out resonance and damps out transition probability. The number of resonance peaks matches the mode number $k$ that defines the excited state of the field. As the acceleration decreases, the resonance peaks becomes narrower and higher, indicating that we approach resonance in static inertial scenario. Figure~\ref{plotresonance1} also shows that resonance dominates when
    \begin{equation}
    \begin{aligned}
        \Omega\tau \approx \lim_{a\to 0} \omega_k t(\tau)\,,
    \end{aligned}
    \end{equation}
    where $t(\tau)$ is the pullback of the coordinate time in terms of proper time $\tau$ of the detector. Crucially, the rough estimate of the right hand side gives
    \begin{align}
    \label{eq:nonrelativistic}
         \omega_k t(\tau)\approx \frac{k\pi}{aL}\rr{a\tau+O(a^3\tau^3)} \sim \frac{k\pi \tau}{L}+O(a^3\tau^3)\,,
    \end{align}
    which is to first order the same as the case for static detector and static cavity. 
    
 We remark that near resonance $\Omega - \omega_k\approx 0$, Figure~\ref{plotresonance1} seem to indicate that the probability amplitude may be divergent if $a$ is small enough, since $\lambda^2$ may not be small enough to make the probability amplitude less than 1 (e.g. set $\lambda=0.1$). We expect this to be an artifact of the approximations in the whole setup, including perturbative calculations of transition probability $P(\Omega)$. As an example of such artifacts, note that in Eq.~\eqref{resonantprob} the transition probability scales linearly with $n_k$ (this also appears in \cite{Dragan:2011zz}). Clearly, this cannot be valid for arbitrary $n_k$ since for large enough excitations, the probability can be made greater than 1. These may be cancelled by higher order terms which would also contain co-rotating terms. Also, recall that since our detector starts from one end of the cavity, in the limit where $a=0$ we should expect \textit{no excitation} at all due to Dirichlet boundary condition given the choice of coupling. This suggests that for computations involving non-vacuum contributions and co-rotating terms, one should be careful in extrapolating results.

    Nonetheless,  our results so far do not change even if we stay away from the $\Omega\approx \omega_k$ limit (cf. Figure~\ref{plotsinglecoherent1}), since all that the resonance condition and large $n_k$ limit do is  allow us to ignore   vacuum contributions from $W_0(\tau,\tau')$ by amplifying the non-vacuum contributions.  Even if the excited parts  $W_{\text{exc}},W_{c}$ of the Wightman function are smaller than $W_0$, we could simply subtract off the $W_0$ part since we find a negligible difference between the responses in Experiment 1 and Experiment 2.

    \section{Transition rate}\label{sec:rate}
    
   Computation of a transition probability --- also known as a \textit{response function} $F(\Omega)$ of a particle detector with energy gap $\Omega$ --- as a detector traverses through a quantum field coupled to it has a  physical interpretation: it provides an operational way of defining the particle content of the field without invoking a high degree of spacetime symmetry \cite{Unruh:1976,DeWitt1979}. However, the fact that it is a double integral may obscure information about the atom-field interaction. This prompts us to consider whether the \textit{transition rate}, essentially the time-derivative  of the response function along the detector trajectory, can provide further insights into the WEP.
    
    To obtain the response rate, we need to rewrite the response function in such a way that it can be easily differentiated. This is done by changing   variables  \cite{Louko:2007mu}
    \begin{align}
        F(\Omega) = 2\text{Re}\int_{\tau_0}^\tau\dd u\int_0^{u-\tau_0}\dd s e^{-i\Omega s}W(u,u-s)\,,
    \end{align}
    where $\tau_0$ denotes the time in which the detector is switched on. Instead of the usual response function which gives transition probability of exciting the atom from its ground state, we can now compute the \textit{instantaneous transition rate} of a detector turned on at time $\tau_0$ and read at time $\tau$, given by \cite{Louko:2007mu}
    \begin{equation}
        \dot F(\Omega) = \frac{\dd F(\Omega)}{\dd\tau} = 2\text{Re}\int_0^{\tau-\tau_0}\dd s\, e^{-i\Omega s}W(\tau,\tau-s)\,.
    \end{equation}
    Despite some subtleties in handling this observable for free space involving regularization, we expect that cavity setup removes these difficulties since the field is compactly supported and there is an infrared cutoff. In our scenario it is convenient to compute the case where $\tau_0=0$. If different field states have a chance of causing different responses to the detector, transition rate may be able to pick this up\footnote{On the other hand, it is possible that response function washes out differential differences due to mean value theorem.}. Conversely, if transition rate is identical, then the response of the detector should be the same under integration.
    
\begin{figure}[htp]
        \centering
        \includegraphics[scale=0.8]{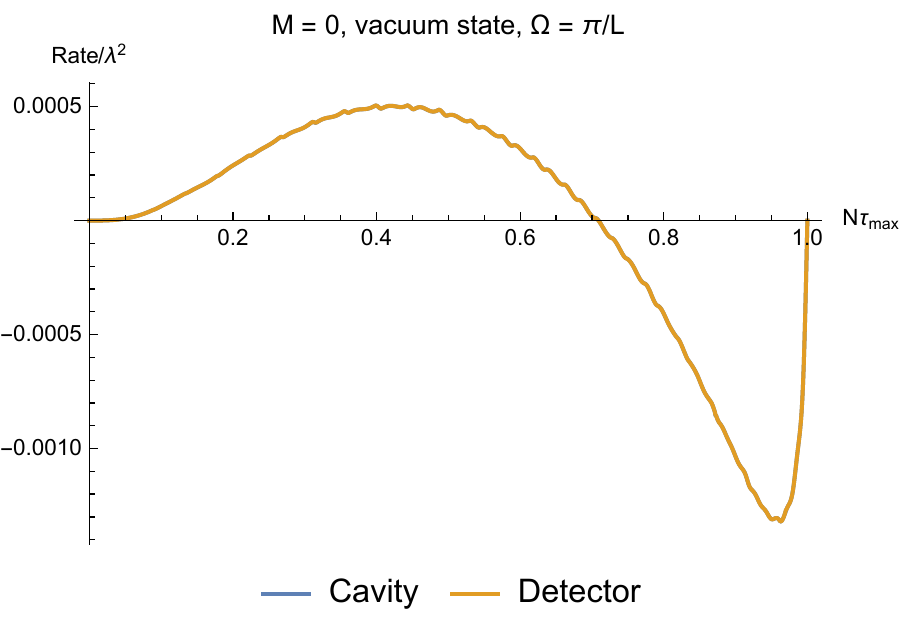}
        \includegraphics[scale=0.8]{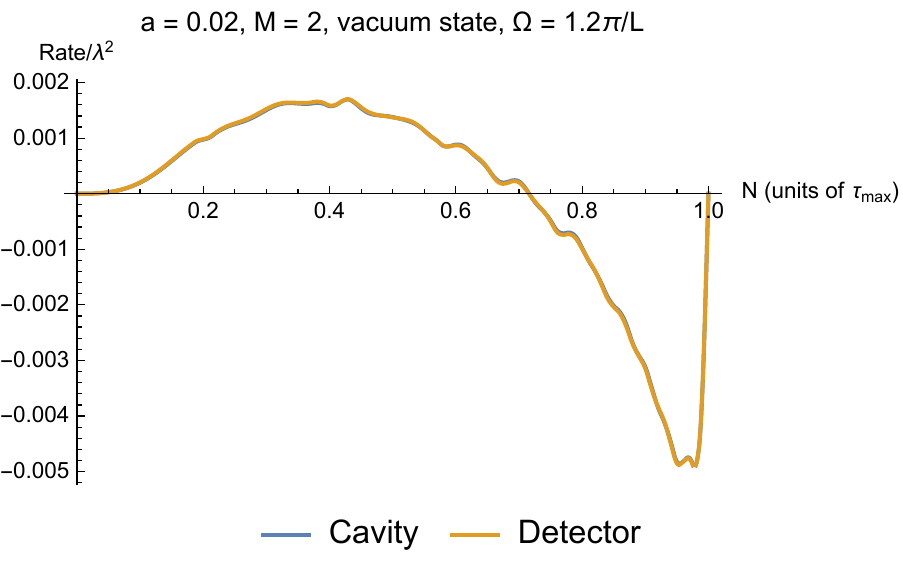}
        \caption{Transition rate as a function of time $\tau$. Note that in both cases the transition rates for accelerating cavity and accelerating detector scenarios are practically indistinguishable regardless of mass. We chose different parameters for variations.}
        \label{plotrate1}
    \end{figure}

    Since the response rate is linear in $W(\tau,\tau')$, we will split them into two parts:
    \begin{equation}
        \dot F(\Omega) = \dot F_0(\Omega)+\dot F_{1}(\Omega)
    \end{equation}
    where $\dot F_0$ is the vacuum contribution and $\dot F_{1}$ is the remaining contribution due to the field in excited state. The vacuum state transition rate is shown in Figure~\ref{plotrate1}. The crucial thing to note here is that the vacuum contribution for both cases have negligible differences in transition rate --- therefore the transition probability must be the same as well after integrating across the full trajectory. For computational time convenience, we chose $a=0.02$ to represent massive case and the same conclusion holds. This justifies our earlier results (also in \cite{Dragan:2011zz}) that vacuum contributions are not sensitive to local accelerations.
    
\begin{figure}[htp]
        \centering
        \includegraphics[scale=0.8]{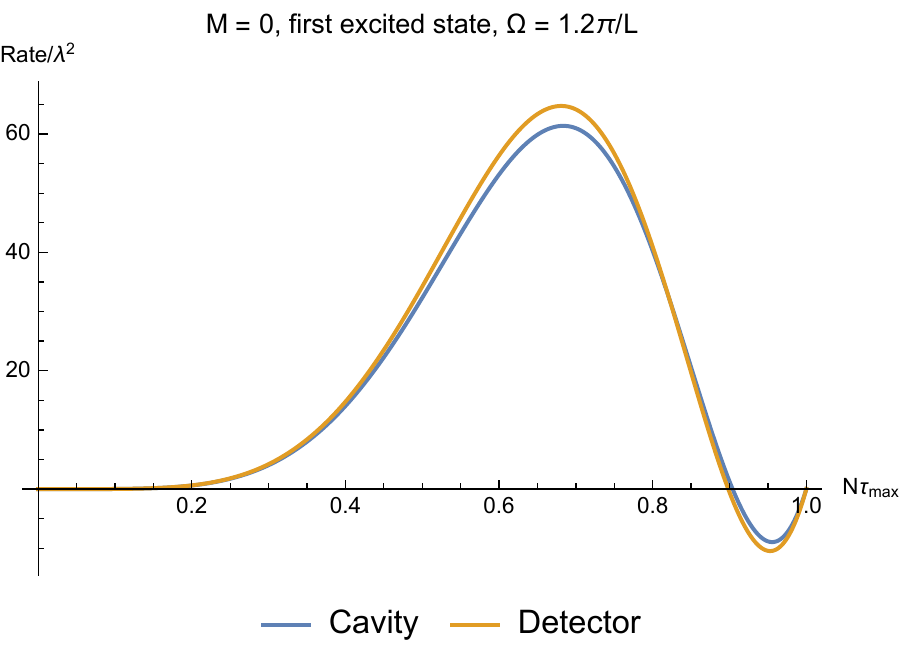}
        \includegraphics[scale=0.8]{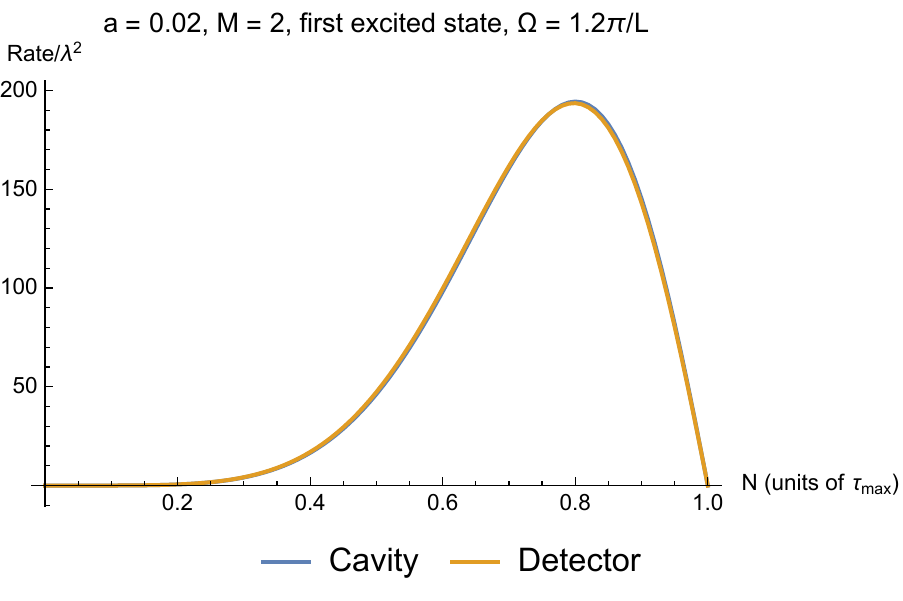}
        \includegraphics[scale=0.8]{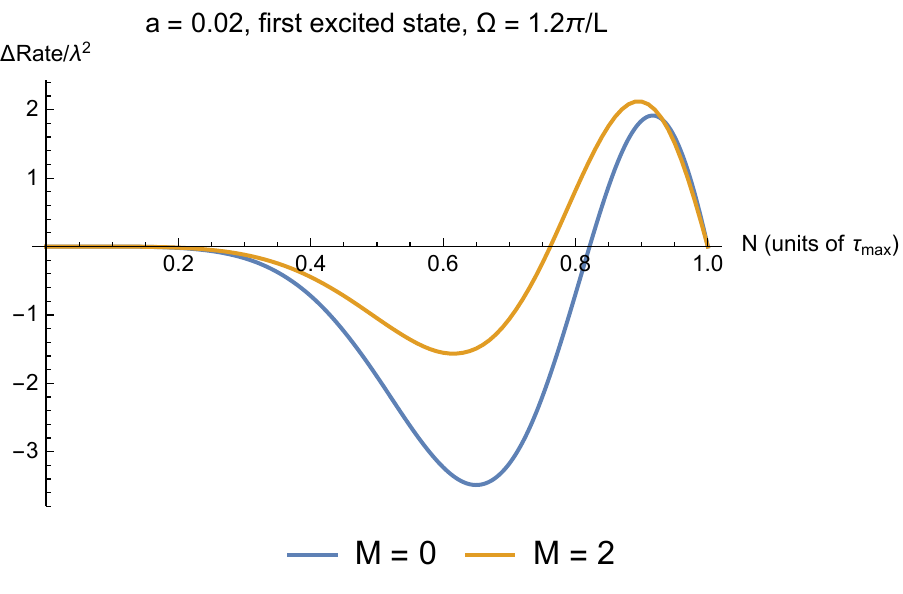}
        \caption{Transition rate as a function of time $\tau$ for the first excited state of the field. \textbf{Top:} massless case. \textbf{Middle:} massive case. \textbf{Bottom:} Difference in transition rate for both scenarios. It appears that transition rate and hence transition amplitude is slightly more advantageous for massless case for a given acceleration. Here ``$\Delta\text{Rate}$" is simply $\Delta\dot F = \dot F^D=\dot F^C$.}
        \label{plotrate2}
    \end{figure}

    Two examples for a highly populated first excited state $(k = 1, n_k = 100)$ for the massless case are shown in Figure~\ref{plotrate2}. We see that while the rate appears qualitatively different at different read-out times, the \textit{difference} between an accelerating cavity and an accelerating detector is very small (of course it is only exactly zero for a static setup). As far as differences go, massive fields generically do not perform better than massless ones, which is consistent with the idea that the role of mass tends to `kill off' correlations at large distances and diminish the amplitudes. 
    
\begin{figure}[htp]
        \centering
        \includegraphics[scale=0.8]{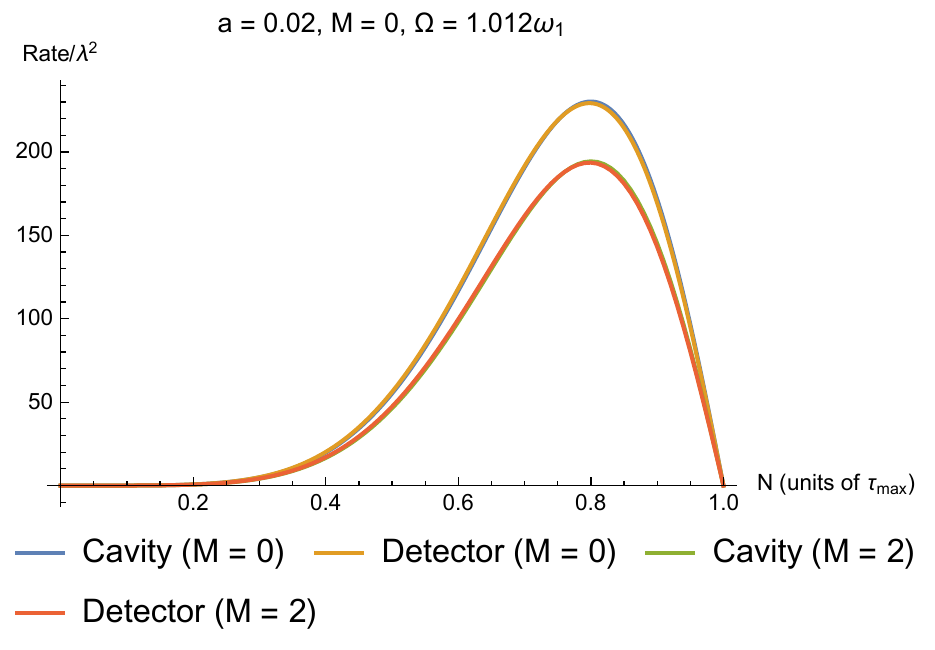}
        \caption{Transition rate as a function of time $\tau$ for the first excited state of the field. \textbf{Top:} massless case. \textbf{Middle:} massive case. \textbf{Bottom:} Difference in transition rate for both scenarios. It appears that transition rate and hence transition amplitude is slightly more advantageous for massless case for a given acceleration.}
        \label{plotrate3}
    \end{figure}

    From Figure~\ref{plotrate2}, one might be led to think that a massive field seems to have a \textit{very large} response rate compared to a massless field, but this is not the right comparison. Note that the co-rotating frequency $\Omega-\omega_n$ determines quite directly the magnitude of these rates, and given the same  gap $\Omega$, one of the two fields will be closer to `resonant frequency' than the other. In Figure~\ref{plotrate3}, we adjust the gap so that both massless and massive fields the atomic gap is $\Omega = 1.012\omega_1$, where $\omega_1$ is the frequency of the first mode. As expected, the absolute value of the transition rate for massless fields dominate the massive case. The difference in response rates $\Delta\dot{F} = \dot F^D-\dot F^C$, where $C,D$ denotes accelerating cavity and detector respectively, are of approximately the same order as seen in Figure~\ref{plotrate3}.
    
    With hindsight we should not be surprised by these results, since they are basically an Unruh-type setup confined to cavity. As clarified in \cite{Fulling:2018lez}, what is important in these WEP considerations is really the fact that there is \textit{relative acceleration} between the atom and the cavity. In the slow acceleration limit, every point in the cavity can be approximated to have the same constant proper acceleration (hence the same clock ticking rates) and so an accelerating cavity-static detector and an accelerating detector-static cavity should lead to the same physical results. The mass parameter of the scalar field enters the quantum field via the mode frequency and amplitude, which generally degrades response since the integral over Wightman function is more oscillatory and the normalization for each mode is smaller than those for a massless field.  In this respect, if a `fair' comparison is made between the massless and massive cases (e.g.  adjusting $\Omega/\omega_k$ or $|\Omega-\omega_k|$ instead of fixing $\Omega$l, cf. Section~\ref{sec:excited}), the massless field should lead to larger detector responses because mass suppresses nonlocal correlations. Note that this suppression is \textit{independent} of WEP.
    
    Why would the responses be different at large $a$? As argued in the context of mirrors \cite{Fulling:2018lez}, the accelerating cavity-static detector and the accelerating detector-static cavity setups are also not mathematically equivalent: if our experiments are sensitive enough to non-uniformity of acceleration across the cavity, then the notion of ``relative acceleration" becomes blurred. For an accelerating detector, in the cavity frame one observes that the detector has a constant-acceleration trajectory; for an accelerating cavity, in the cavity frame observes that the detector is not uniformly accelerating because its worldline crosses all the hypersurfaces of constant $\xi$ between one cavity wall $\xi=\xi_2$ to another $\xi=\xi_1$. In the slow acceleration limit, these constant-$\xi$ surfaces describe approximately the same acceleration and hence the detector is observed to be approximately uniformly accelerating. We can think of the correlation functions of the field as capturing this non-local difference and the inequivalent setups lead to unequal responses. It is in this spirit that WEP makes sense --- the responses between the free-falling cavity-stationary detector and free-falling detector-stationary cavity will be different once the non-uniformity of the gravitational field is detectable.
    
   Finally, a small qualification about the comparison between the two different scenarios (accelerating detector and accelerating cavity) is in order. There are a couple of ways in which the two scenarios can be argued not to be on equal footing, First, we note that in relativity there is no absolute rigidity \cite{Herglotz1910,Noether1910,Epp:2008kk}; it is impossible to maintain fixed coordinate distance between two cavity walls in \textit{all frames}. Accelerating the cavity whilst keeping it  rigid in the cavity rest frame (Fermi-Walker rigidity) is  the simplest and most natural setup. The fact that for accelerating cavities the detector is seen to be non-uniformly accelerating from the cavity's frame, is sufficient to show that the detector response should be different from the constantly accelerating detector scenario. 
   
   However, there is also a perhaps more fundamental and easier argument for the lack of equivalence between the two scenarios: The accelerating cavity is a setup of accelerating mirrors, which are perfectly reflecting boundary conditions, whereas  an accelerating detector is a quantum object that can absorb, transmit and reflect parts of an illuminating plane wave. Consequently, they constitute rather distinct field configurations (e.g., dynamical Casimir effect and Unruh radiation respectively) and the two setups are not identical beyond the `non-uniformity' of the acceleration either. We can estimate the deviation between the two scenarios e.g. from Eq.~\eqref{eq:nonrelativistic}, which can be seen to be third order in the dimensionless parameter that depends on acceleration and duration of the interaction.

    \section{Conclusion}
    
 We have investigated a quantum version of the WEP in which we consider the response of a particle detector in two scenarios: a) a detector accelerating in a static cavity  and b) a static detector in an accelerating cavity. We found that the qualitative WEP is indeed satisfied insofar as quasilocal approximations are valid. We do this by investigating the transition probability of a two-level atomic detector on various field states, namely vacuum state (Minkowski-like and Rindler-like vacuum), arbitrary Fock state, and single-mode coherent state. We also check the effect of bringing the atomic gap closer to the resonant frequency when we have co-rotating terms and clarify the validity of some approximations such as large $n_k$ limit for Fock state of the field.  Importantly, the results support the idea that a `quantum accelerometer' in non-relativistic regime would work equally for a massless field and for a massive field. We strengthen the results by computing the transition rates to ensure that no fundamental physical differences are averaged out by integration when we compute transition probabilities.  In this sense, our results complement those of \cite{Scully:2017utk,Fulling:2018lez}.
    

    \section*{Acknowledgement}
    This work was supported in part by the Natural Sciences and Engineering Research Council of Canada. E. Tjoa thanks Jorma Louko, Robie Hennigar and Richard Lopp for useful discussions and A. Dragan for helpful correspondence. E. M-M also acknowledges funding form his Ontario Early Researcher Award.
    
    
    \appendix
    \section{Solving massless Klein-Gordon equation without conformal transformation}
    \label{appendix:conformal}
    
    In this section we solve for the solution for the massless Klein-Gordon field equation without invoking conformal transformation of any sort. We quote again the standard Rindler coordinates for convenience:
    \begin{align*}
        t=\xi\sinh\eta\,,\hspace{0.5cm}x=\xi\cosh\eta\,.
    \end{align*}
    From the general Klein-Gordon field equation (cf. Eq.~\eqref{kleingordon}) in this coordinate system, which gives the modified Bessel differential equation for the spatial modes $v(\eta,\xi)$:
    \begin{align}
    \label{modifiedbesselODE}
        \xi^2\frac{\dd^2 v}{\dd \xi^2}+\xi\frac{\dd v}{\dd\xi} +(\omega^2-m^2\xi^2)v = 0\,.
    \end{align}
    The solution basis for $m\neq 0$ is given by $\text{Re}(I_{i\omega})$ and $K_{i\omega}$ which are both real and linearly independent due to nontrivial Wronskian \cite{NIST:DLMF}. Now let us set $m=0$ on Eq.~\eqref{modifiedbesselODE}.
    The eigenbasis\footnote{This is not the ones used in e.g. \cite{Friis:2013a,Dragan:2011zz}, but for our purposes either one will work. Roughly speaking, one can check from the series expansion at small $m$ that this is analogous to the choice of writing solutions to harmonic oscillator equation in terms of cosine/sine functions or plane waves.} of the solution space is given by $\sin \rr{\omega \log \xi}$ and $\cos \rr{\omega \log \xi}$. Note that we could also obtain this by doing a series expansion for small $m\to 0^+$ on the mode solutions in Eq.~\eqref{Besselbasis} which satisfies the Dirichlet boundary condition at $\xi=\xi_1$ 
    \cite{NIST:DLMF}. Since $\eta$ is dimensionless, so is $\omega$ here. If we let the boundary conditions to be at $\xi_1=a^{-1}$ and $\xi_2=a^{-1}+L$, we get
    \begin{equation}
        v_n \propto \sin \rr{\omega_n\log \xi}-\tan\rr{\omega_n\log \frac{1}{a}}\cos\rr{\omega_n\log \xi}\,,
    \end{equation}
    where $\omega_n$ is now a discrete spectrum due to the second boundary condition $\xi=\xi_2$. The normalization can be found by standard Klein-Gordon inner product \cite{birrell1984quantum}. Remarkably, even after imposing the second boundary condition, the spectrum is still exact, which reads
    \begin{equation}
        \omega_n = \frac{n\pi}{\log(1+aL)}\,,\hspace{0.5cm}n\in \mathbb{N}\,,
    \end{equation}
    which is precisely what we got from the conformal transformation where we identify the denominator as $aL'$, the conformally transformed length of the cavity multipiled by the kinematical parameter $a$. In some sense this is perhaps not surprising, since the same physical situation should be described by the same differential operator with the same set of spectrum (which is invariant under coordinate transformations). 
    
\begin{figure}[htp]
        \centering
        \includegraphics[scale=0.8]{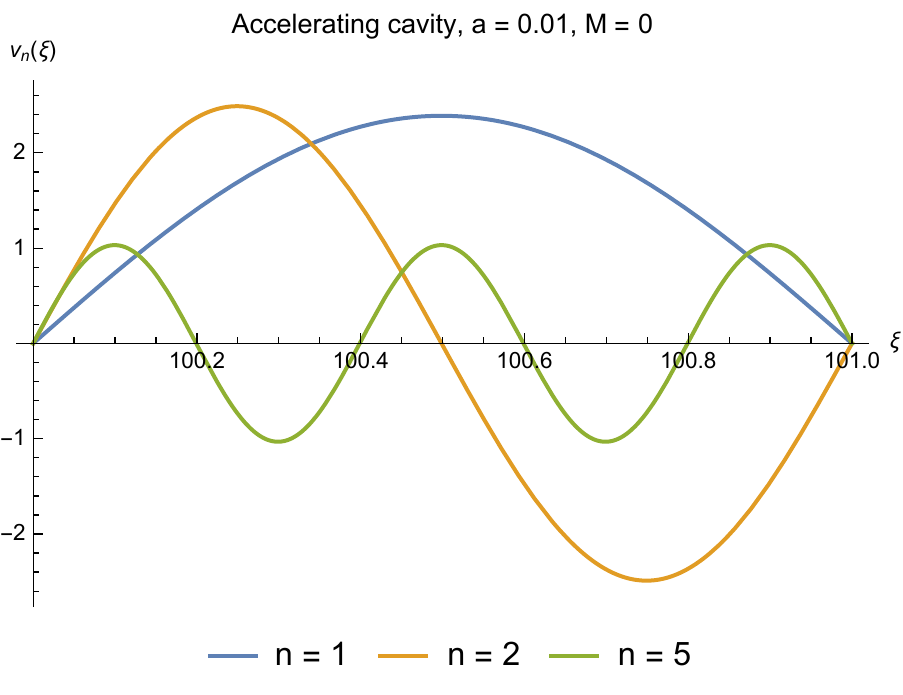}
        \includegraphics[scale=0.8]{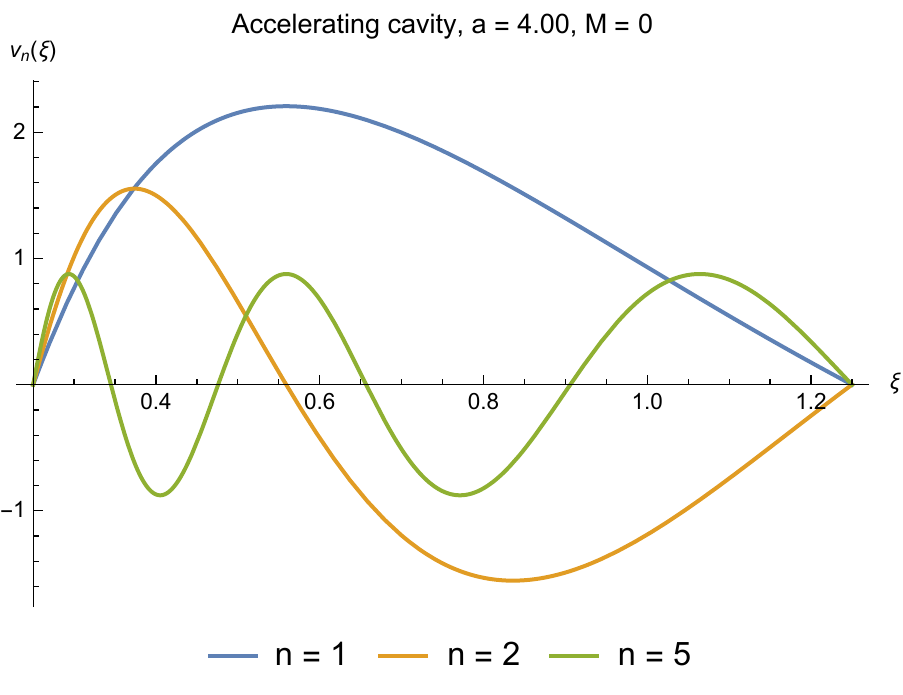}
        \caption{Sample plots of mode functions for the second mode $n=2$ for small and large accelerations. This makes clear that large acceleration limit is ``Bessel-like", in that the mode function is a deformed sine function, squashed in the direction of acceleration. These plots are not normalized since we are concerned with their forms rather than their amplitudes.}
        \label{plotconformal1}
    \end{figure}

    Some representative plots of the modes for small and large accelerations are given in Figure~\ref{plotconformal1}. Now it is very clear that the spatial modes approach Minkowski static cavity scenario very quickly for not too small $a\sim 0.01$, while for large acceleration (of the left wall) the modes are ``deformed sine functions". These deformed modes are in fact very similar in form as the modes for massive case described in terms of modified Bessel functions of imaginary order $\text{Re}(I_{i\omega})$ and $K_{i\omega}$. 
    
    This clearly demonstrates that the differential equation governing the form of the spatial modes is solvable directly even if the metric is not the one conformally equivalent to the Minkowski metric. In this standard Rindler coordinates, the Klein-Gordon equation would also not be conformally invariant under the change of coordinates. However, the standard Rindler coordinates and conformal Rindler coordinates both cover the Rindler wedge portion of Minkowski spacetime and each hypersurface of constant $\xi$ in either coordinates describe the trajectory of uniformly accelerating test particles. One would not conclude that massless fields cannot distinguish the two scenarios on grounds of conformal invariance, while massive fields can; instead, one would conclude that both should have qualitatively similar behaviour up to some degradation factor due to mass of the field that enters the normalization constant and phase factor in the integral of transition probability.
    
    Here we make a short remark on the distinction between conformal flatness and conformal invariance of field equations via a conformal transformation. A spacetime $M$ is said to be conformally flat if \textit{there exists} a coordinate system in which the metric can be rewritten as 
    \begin{equation}
        g_{\mu\nu}(\sx) = \Omega(\sx)^2\eta_{\mu\nu}\,,
    \end{equation}
    and in (1+1) dimensions all Lorentzian manifolds are conformally flat. The massless KG field is conformally invariant because under conformal transformation, the KG equation takes the same form as the wave equation in global Minkowski coordinates. However, performing conformal transformation is a calculational advantage that does not change the physics, since we could equally do physics using non-conformally equivalent metric that describes the same spacetime. Alternatively, we say that the physics is contained in $\Omega(\sx)$ and so the physics will still be different from static Minkowski spacetime \cite{Fulling:2018lez}. A good example is the de Sitter expanding universe, which can be written in coordinates such that it is conformally flat --- the mode functions inherit the form in flat space, but static detector in conformal vacuum of the de Sitter spacetime detects particles while static detector in Minkowski vacuum does not. 
    
     \begin{figure}[H]
        \centering
        \includegraphics[scale=0.7]{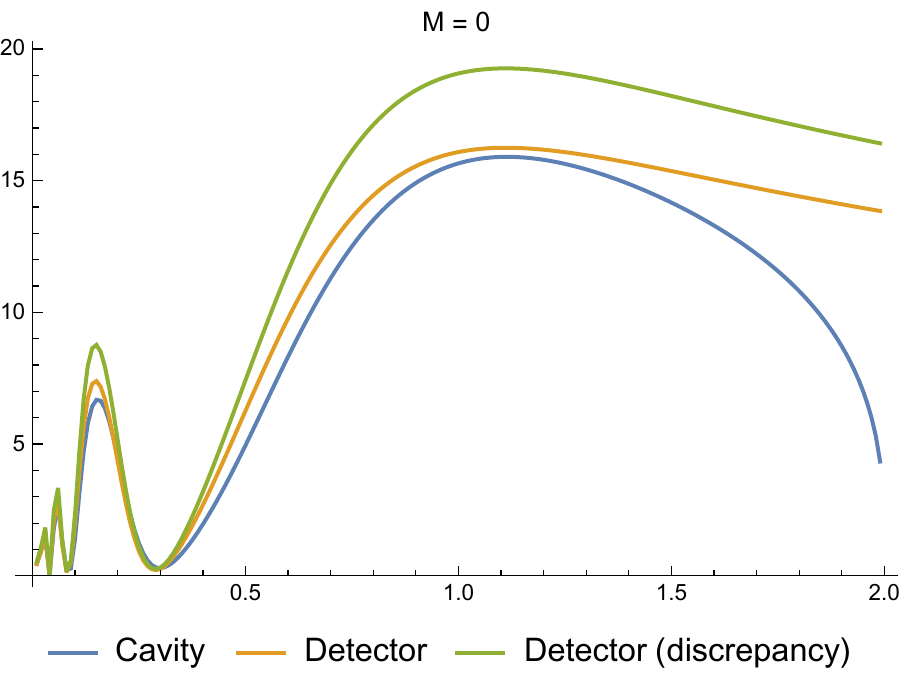}
        \includegraphics[scale=0.7]{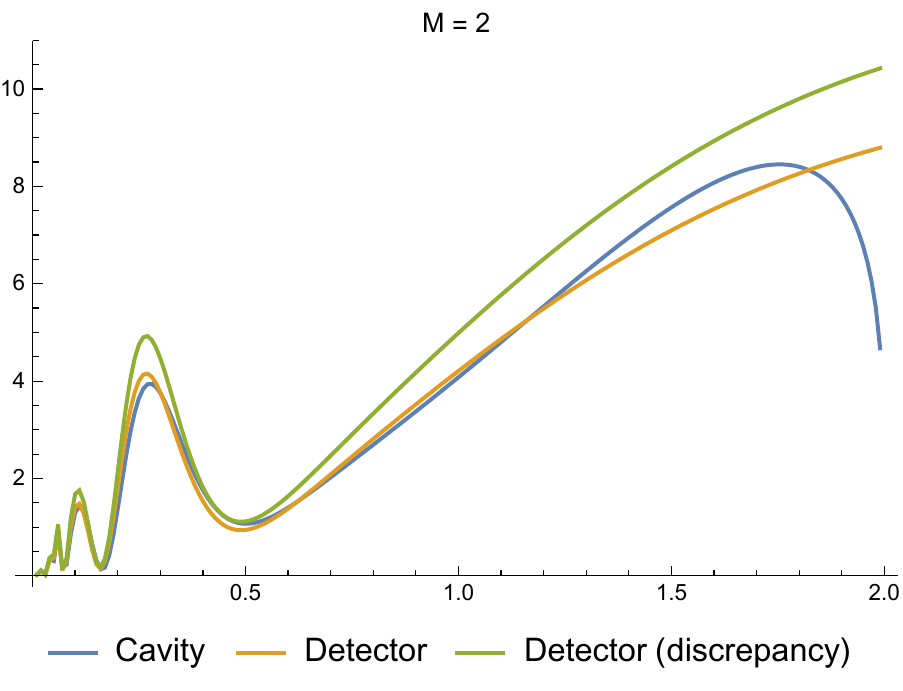}
        \caption{The transition probability plot simulating the plots found in \cite{Dragan:2011zz}. The discrepancy is possibly related to incorrect normalization for massive field-accelerating detector scenario (labelled `discrepancy' here), thus producing the result that massive fields can better distinguish local acceleration.}
        \label{plotwrong}
    \end{figure}

    \section{Discrepancy with past results}
    \label{appendix:disparity}
    Based on the argument above, there is a slight disparity in a result we obtained here and the results obtained in \cite{Dragan:2011zz}. Since the exact parameters used previously \cite{Dragan:2011zz} are unknown, we attempt to emulate the construction and the result is shown in Figure~\ref{plotwrong}. From what we can discern, this discrepancy arises from making the same  (inappropriate) normalization choice for both the massive
    and massless cases.  For a   detector accelerating in a static cavity with a massive scalar field \cite{Dragan:2011zz}, this leads to the conclusion that (in the non-relativistic regime) massive fields can distinguish local acceleration whereas massless fields cannot.

    Despite the discrepancy, the results here and in \cite{Dragan:2011zz} nonetheless show that detector responses can indeed detect non-uniformity of accelerations in cavity which lead to distinguishability between the two scenarios. Essentially, it boils down to the fact that in the accelerating cavity scenario, the static detector is only approximately uniformly accelerating from the perspective of cavity frame, since the vertical worldlines cross hypersurfaces of constant but different $\xi$, which is approximately constant for very short cavity or very small accelerations. On the other hand, an accelerating detector is an \textit{exactly} uniformly accelerating test body;  thus the setup is not mathematically equivalent --- hence ``qualitative weak equivalence principle" \cite{Fulling:2018lez}.
    
 To summarize, we first note that both accelerating cavity and accelerating detector setups are \textit{kinematically inequivalent} for any nonzero $aL$, as illustrated in Section~\ref{sec:cavity} and Appendix~\ref{appendix:conformal}. What conformal invariance in $(1+1)$ dimensions gives us is convenience, a point made also in \cite{Fulling:2018lez}. It boils down to the fact that  in the rest frame of an accelerating cavity the detector does \textit{not} undergo uniform acceleration. Therefore, for any value of $aL$, there exists a finite difference in transition probability $\Delta\Pr = \abs{\Pr_{cav}-\Pr_{det}}$ between the two setups regardless of the mass of the field. This difference quickly vanishes as $aL \to 0$: in this `quasilocal regime',  we can approximate the whole cavity as accelerating with a single proper acceleration, recalling that the acceleration along the length of the cavity $a(x)$ is related to the acceleration of the rear wall $a_1$ by
    \begin{align}
        a(x) = \frac{a_1}{1+a_1(x-x_1)}\approx a_1
    \end{align}
    if $a_1(x-x_1)<a_1 L \ll 1$. For this reason, $\Delta \Pr$ falls quickly as $a\to 0$, becoming exactly zero when $a=0$ (entirely static detector and cavity setups). So long as $aL\neq 0$, in principle we can always distinguish local accelerations using nonlocal correlations of the field regardless of mass. Choosing the detector gap to be closer to the resonant frequency of the field (e.g. excited Fock state) will help in amplifying very small transition probabilities, noting that the resonant frequencies between massless and massive cases would be different.
  
 An alternative interpretation would be to require  that if $\Delta \Pr$ is below certain threshold we lose the capacity to distinguish local accelerations in the non-relativistic regime. All things being equal (taking into account resonant effects etc.), this would mean that generically neither massless nor massive fields can do the job 
 of frame distinction if the threshold is not exceeded. While operationally sensible, we prefer the previous interpretation since $\Delta \Pr$ generally never actually vanishes except when both the cavity and the detector are at rest relative to one another. Neither massless nor massive fields are `preferred' in their capacity to distinguish local relative accelerations; any quantitative difference is purely due to quantum-theoretic aspects of  nonlocal field correlations and their dependence on mass.

    \section{Convergence of mode sums}
    \label{appendix:convergence}
\begin{figure}[H]
        \centering
        \includegraphics[scale=0.9]{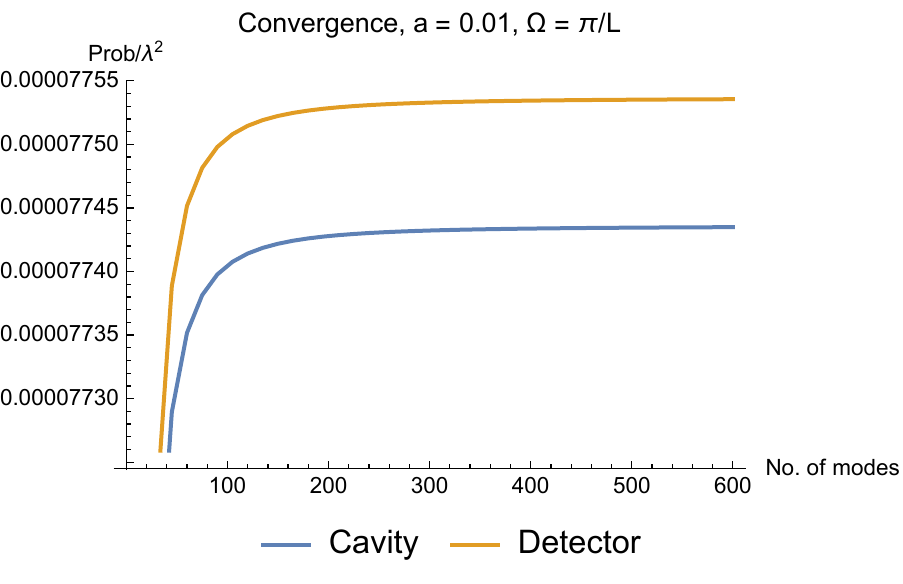}
        \includegraphics[scale=0.9]{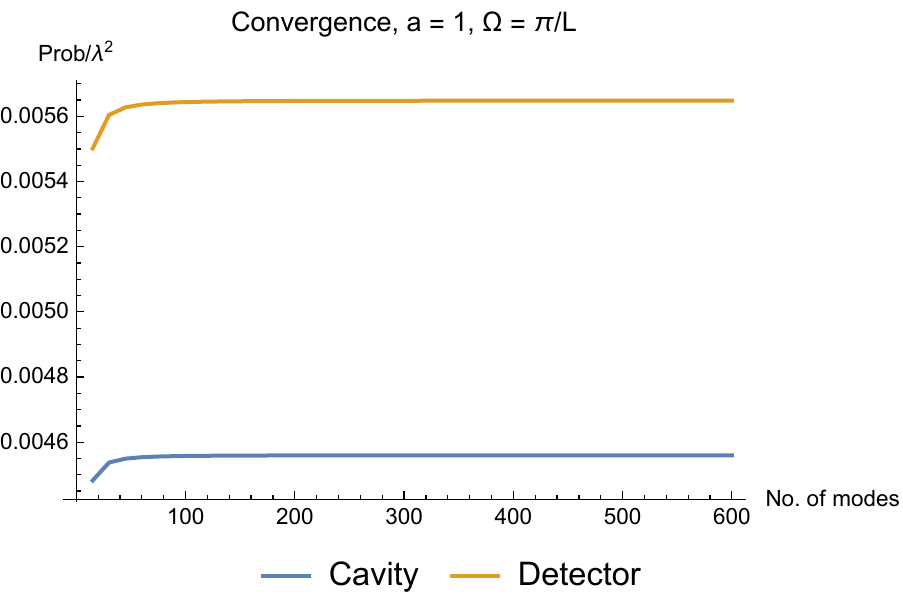}
        \caption{Probability as a function of mode sum $N$ for $a=0.01$ and $a=1.0$ with $M=0$.}
        \label{cavconvergence1}
    \end{figure}

    We show some plots demonstrating how quickly the mode sums converge for certain choices of parameters. In Figure~\ref{cavconvergence1} we plot the transition probabilities
    as a function of mode sum for the field initiated as vacuum state for two different accelerations $a$.
    
    We see that the convergence is attained for relatively small $N\sim 100$, and even if we sum $N = 15$ (the smallest $N$ in these plots), the values do not stray far from the converged value, thus for practical purposes we choose to perform calculations involving vacuum state for $N=15$. Note that for fields initiated in excited states, the Wightman function has vacuum and excited state contributions but the latter does not occur as sums over modes and hence convergence issue does not appear.

    \bibliography{myref}

\end{document}